\documentclass[aps,twocolumn,preprintnumbers,amsmath,amssymb,superscriptaddress]{revtex4-1}

\usepackage{amsmath,cleveref,amssymb,amsfonts,graphicx,multirow,color,bbm,textcomp,mathtools,dsfont}
\usepackage{paralist}
\usepackage{srcltx}
\usepackage[all,cmtip]{xy}
\usepackage{mathrsfs}
\usepackage{srcltx,soul,subfigure}
\usepackage{txfonts}

\newcommand{\ket}[1]{| #1 \rangle}

\newcommand{\bee}{\begin{equation}}
\newcommand{\ee}{\end{equation}}
\newcommand{\bma}{\begin{pmatrix}}
\newcommand{\ema}{\end{pmatrix}}
\newcommand{\balig}{\begin{align}}
\newcommand{\ealig}{\end{align}}

\newcommand{\ba}{\begin{align}}
\newcommand{\ea}{\end{align}}

\newcommand{\ignore}[1]{}

\newcommand{\six}{\sigma_x}
\newcommand{\siy}{\sigma_y}
\newcommand{\siz}{\sigma_z}

\newcommand{\br}{\bm{r}}

\newcommand{\sgn}[1]{\signfunction{(#1)}}

\usepackage{siunitx}

\newcommand{\average}[1]{\langle#1\rangle}
\newcommand{\TwoDMatrix}[4]{\begin{pmatrix} #1 & #2\\#3 & #4\end{pmatrix}}

\newcommand{\abs}[1]{\left|#1\right|}

\newcommand{\brackets}[1]{\left[#1\right]}
\newcommand{\parenthesis}[1]{\left(#1\right)}

\newcommand{\bI}{\mathbbm{1}}

\def\sgn{\mathop{\textrm{sgn}}}

\newcommand{\bk}{{\bf k}}

\begin{document}


\title{Type-II Dirac surface states in topological crystalline insulators}

\author{Ching-Kai Chiu}
\affiliation{Department of Physics and Astronomy, University of British Columbia, Vancouver, BC, Canada V6T 1Z1} 
\affiliation{Quantum Matter Institute, University of British Columbia, Vancouver BC, Canada V6T 1Z4}
\affiliation{Condensed Matter Theory Center and Joint Quantum Institute and Maryland Q Station, Department of Physics, University of Maryland, College Park, MD 20742, USA}

\author{Y.-H. Chan}
\affiliation{Institute of Atomic and Molecular Sciences, Academia Sinica, Taipei 10617, Taiwan}

\author{Xiao Li}
\affiliation{Condensed Matter Theory Center and Joint Quantum Institute and Maryland Q Station, Department of Physics, University of Maryland, College Park, MD 20742, USA}

\author{Y. Nohara}
\affiliation{Max-Planck-Institute for Solid State Research, Heisenbergstr. 1, D-70569 Stuttgart, Germany}


\author{A. P. Schnyder}
\affiliation{Max-Planck-Institute for Solid State Research, Heisenbergstr. 1, D-70569 Stuttgart, Germany}

\date{\today}

\begin{abstract} 
We study the properties of a family of anti-pervoskite materials,
which are topological crystalline insulators with an insulating bulk but a conducting surface.
Using   ab-initio DFT calculations, we investigate  
the bulk and surface topology and show that these materials exhibit type-I as well as
type-II Dirac surface states protected by reflection symmetry. 
While type-I Dirac states give rise to closed circular Fermi surfaces, type-II Dirac surface states
are characterized by open electron and hole pockets that touch each other.
We find that the type-II Dirac states  exhibit characteristic van-Hove singularities in their dispersion,
which can serve as an experimental fingerprint.
In addition, we study the response of the surface states
to magnetic fields.
%
%
%

\end{abstract}

\maketitle

\noindent
Topological crystalline insulators (TCIs) are insulating in the bulk, but exhibit 
conducting boundary states protected by crystal 
symmetries~\cite{Ando_Fu_TCI_review,Teo:2008fk,chiu_RMP_16}. 
As opposed to surface states of ordinary insulators, the gapless modes at the surface of TCIs arise due to a non-trivial  topology of the bulk wavefunctions, which is characterized by a quantized topological invariant, e.g., a mirror Chern or mirror winding number~\cite{Chiu_reflection,Morimoto2013,Sato_Crystalline_PRB14}.
One prominent  example of  a TCI is the rocksalt semiconductor SnTe~\cite{Dziawa:2012uq,Hsieh:2012fk,Tanaka:2012fk,Xu:2012},
which supports at its (001) surface four Dirac cones protected by reflection symmetries.
The surface modes of this  and all other known TCIs are of the
standard Dirac fermion type with closed point-like (or circular) Fermi surfaces, which we refer to as ``type-I".
However, as we discuss in this article, crystal symmetries can also give rise to other types of surface fermions.

In particular, we show that at certain surfaces of the anti-perovskite materials A$_3$EO~\cite{OgataJPSJ2011,kariyadoJPSJ12,fuseyaJPSJ12,kariyadoJPSJ12,KariyadoPhDThesis,Nuss:dk5032,Nuss:dk5032,TCI_Fu_antiperovskites}   there exist so called ``type-II" Dirac points which are protected by reflection symmetry. (Here A denotes an alkaline earth metal, while E stands for Pb or Sn.) 
These type-II band crossings do not lead to circular Fermi surfaces, but
rather give rise to open electron and hole pockets that touch each other. This is reminiscent
of the type-II Weyl points that have recently been predicted to exist in  WTe$_2$~\cite{Soluyanov:2015aa,2016arXiv160401398M,2016arXiv160404218W,2016arXiv160405176W} and LaAlGe~\cite{LaAlGe_II}. 
Here, however, these type-II band crossings occur at the surface rather than in the bulk of the material.
As a consequence, the surface physics of A$_3$EO is radically different to the one of standard TCIs with 
point-like Dirac surface states. This pertains in particular to the surface magneotransport.  
 
Using  first-principles calculations and a tight-binding model,
we systematically study the surface states of A$_3$EO, with a particular focus on Ca$_3$PbO, which
crystallizes in its low-temperature phase in the cubic space group $Pm\bar{3}m$.
We find that the (011) surface exhibits type-II Dirac nodes, whereas the (111) surface supports both type-I and type-II Dirac states.
On the (001) surface, on the other hand, the Dirac nodes overlap with the bulk bands. 
All these surface states are protected by the crystal symmetries of  A$_3$EO,
in particular the reflection symmetries. We show that 
the type-II Dirac nodes exhibit characteristic van Hove singularities in their dispersions,
which can be used as an experimental fingerprint. Moreover, 
we analyze the response of the surface states to a magnetic field 
and show how the Landau level spectra of type-II Dirac nodes
radically differ from those of type-I Dirac states. 
In particular, we find that for type-II Dirac states there is a very large density of Landau levels
near the energy of the Dirac point.
Moreover, as opposed to type-I Dirac states, the type-II Dirac nodes do not exhibit a zeroth Landau level that
is pinned to the energy of the Dirac band crossing.


	




\vspace{0.5cm}

\noindent
\textbf{\large Results}\\
Before considering the detailed topology of the anti-perovskites A$_3$EO, let us first present 
a general discussion of the types of fermions that can arise at the surface of topological (crystalline) insulators.

\noindent
{\bf Types of surface fermions.}
The Hamiltonian describing the low-energy physics of two-dimensional (2D) surface Dirac states is of the generic form
\bee \label{general_surf_ham}
H_{\rm{surf }}(\bk )= \sum_{\substack{ i=1,2 \\ \alpha=0,1,2 }}k_i A_{i\alpha}\sigma_\alpha ,
\ee
where $k_i$ denote the two surface momenta, $\sigma_i$ are the Pauli matrices, and $\sigma_0$ 
represents the $2 \times 2$ identity matrix. 
The energy spectrum of $H_{\rm{surf }}$  is given by 
\bee \label{generic_energy_spectrum}
E_\pm = \sum_i k_i A_{i0} \pm \sqrt{\sum_j \left(\sum_i k_i A_{ij} \right)^2}\equiv L(\bk)\pm W(\bk),
\ee
where the sums are over $i, j=1,2$. The second term $W(\bk)$ in Eq.~\eqref{generic_energy_spectrum}
is the linear spectrum of standard (i.e., type-I) Dirac fermions. 
That is, for $L(\bk)< W(\bk)$, the surface state is categorized as type I.
Since the energy of type-I Dirac states is dispersive in all  directions, we require that  $\det(A_{ij})\neq 0$. 
On the other hand, if there exists one momentum direction $\bk_0$, such  that $L(\bk_0) > W(\bk_0)$,  
we categorize the surface state as \mbox{type II}. This is
in analogy to the three-dimensional (3D) type-II Weyl points, which have been recently discovered in WeTe$_2$~\cite{Soluyanov:2015aa}. 
We note that, while 3D Wely points are stable in the absence of any symmetry (except translation), two-dimensional
surface Dirac states can only exist in the presence of time-reversal symmetry or spatial symmetries, such as reflection.


Now, the interesting question is how these symmetries restrict the form of Eq.~\eqref{general_surf_ham}.
For a 3D  time-reversal (TR) invariant strong topological insulator, the Hamiltonian $H_{\rm{STI}}$ satisfies $T H_{\rm{STI}}(-\bk)T^{-1}= H_{\rm{STI}}(\bk)$,  with the time-reversal operator $T=\sigma_y K $ and  the complex conjugation operator $K$. This symmetry locks the Dirac node at the time-reversal invariant points of the BZ, such as, e.g., $\bk=0$. 
We observe that the  linear term $k_i \sigma_0$ is forbidden by time-reversal symmetry. 
Hence, the surface states of TR invariant strong topological insulators are described by
\bee
H_{\rm{STI}}(\bk)=k_y\sigma_x -k_x \sigma_y ,
\ee
and are therefore always of type I.


However, type-II Dirac fermions can appear at the surface of  reflection symmetric TCIs (and weak TR symmetric topological insulators). 
The reflection symmetric Hamiltonian of these type-II surface states is generically given by
\bee \label{type_two_state}
H_{\rm{TCI}}^{\rm{surf}}(k_x,k_y)=  A k_y\sigma_0 + k_y\sigma_x -k_x \sigma_y  ,
\ee
with $A>1$. (Without loss of generality we have set the Fermi velocities to $1$ and assumed that the reflection plane is $k_x=0$.)
The type-II Dirac state~\eqref{type_two_state} is protected by reflection symmetry $x \to - x$, 
which acts on~\eqref{type_two_state} as 
\bee 
R_x H_{\rm{TCI}}^{\rm{surf}}(-k_x,k_y) R_x^{-1}= H_{\rm{TCI}}^{\rm{surf}}(k_x,k_y) ,
\ee
with the reflection operator $R_x=\sigma_x $.
Since reflection flips the sign of $k_x$, it allows the linear term $A k_y\sigma_0$ but forbids  $k_x \sigma_0$.

The crucial difference between type-I and type-II Dirac surface states is that the former have 
closed circular Fermi surfaces,
whereas the latter exhibit open electron and hole pockets which touch each other.
As one varies the Fermi energy $E_F$, the Fermi surface of type-I Dirac states
can be shrunk to a single point, which is called  a type-I Dirac point.
In contrast, type-II Dirac states give rise to electron and hole pockets, whose size 
depends on the Fermi energy. At a certain $E_F$ the electron and hole Fermi surfaces
touch each other, 
which is called
a type-II Dirac point.
As opposed to type-I Dirac points, the density of states at type-II Dirac points
remains finite. 
In addition, we observe that in type-II Dirac states one of the two surface bands must bend over
in order to connect bulk valence and conduction bands. As a consequence, there is a maximum 
in the dispersion of the surface states. The latter gives rise to a van Hove singularity, which leads
to a kink in the surface density of states. This can be used as an experimental fingerprint of type-II Dirac states.


\noindent


\begin{table}[tbp]
\centering
\begin{ruledtabular}
\begin{tabular}{cclc}
& tol.\ factor &  bulk gap & topology \\ 
\hline 
Ca$_3$PbO 	& 0.999	& $\sim 15$ meV 	& nontrivial \\
Ca$_3$SnO	& 0.993	& $\sim 5$ meV	& nontrivial  \\
Sr$_3$PbO  	& 0.978	& $\sim 18$ meV 			& nontrivial\\
Sr$_3$SnO 	& 0.973	& $\sim 7$ meV  & nontrivial \\
Ba$_3$PbO 	& 0.962	& $\sim 10$ meV  & nontrivial \\
Ba$_3$SnO 	& 0.957	& gapless &  -- \\
\end{tabular}
\end{ruledtabular}
\caption{
Familiy of anti-perovskite materials  with cubic space group Pm$\bar{3}$m.
The tolerance (tol.) factor 
indicates the deviation from the ideal 
inverse perovskite structure~\cite{Nuss:dk5032}. 
The bulk gap values are obtained from ab-initio first principles calculation.
The wavefunction topology is determined by the mirror Chern numbers $C_{x^0}$,  $C_{x^\pi}$, 
and $C_{x,y}$  (see Methods).
We find that the topology is nontrivial, with $C_{x^0}=C_{x,y}=2$ and $C_{x^\pi}=0$, for all compounds except for Ba$_3$SnO. 
\label{tab:anti-perov_family} 
\label{mTab1}
}
\end{table}

\begin{figure*}[t]
\includegraphics[clip,width=1.5\columnwidth]{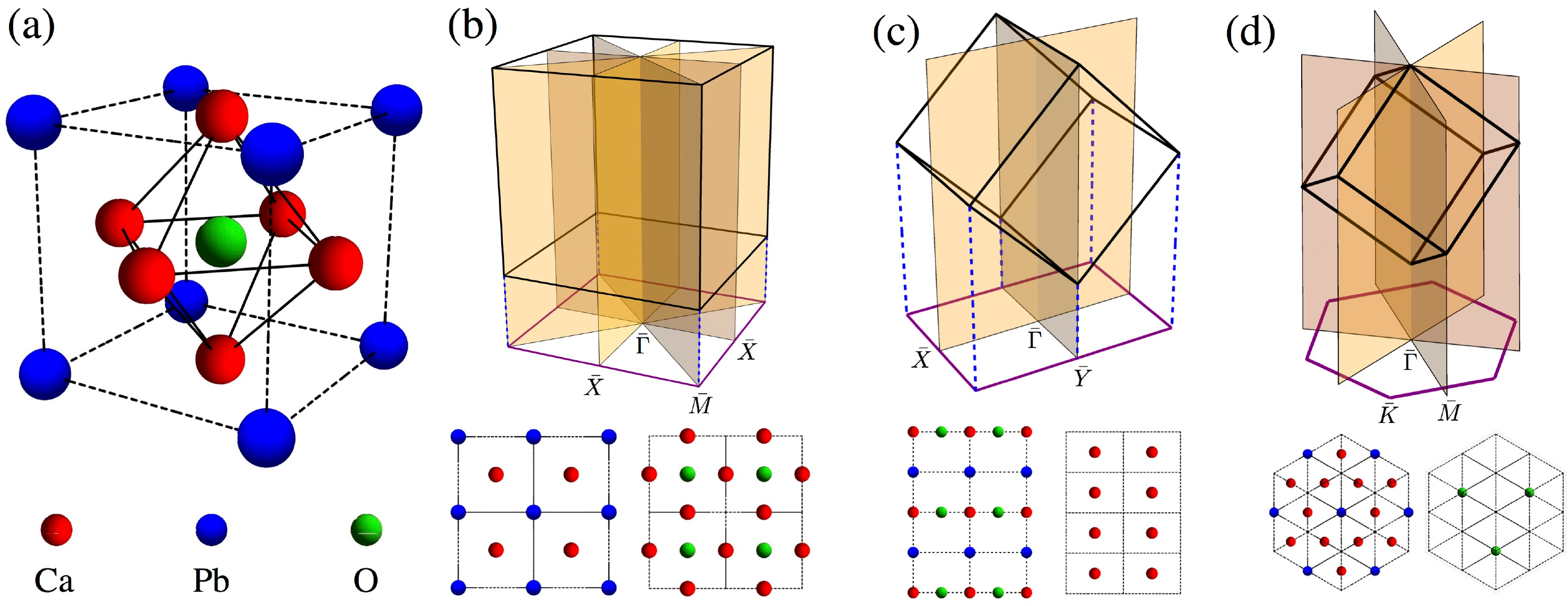}
 \caption{  \label{mFig1}
 {\bf Crystal symmetries and surface terminations.}
(a) Crystal structure of Ca$_3$PbO.
(b), (c) and (d) Surface Brillouin zones (BZs) for the  (001), (011), and (111) surfaces, respectively. 
For the (001) and (011) surfaces the surface BZs can be obtained from the bulk cubic BZs by projecting along 
the [001] and [011] directions, respectively. 
For the (111) surface BZ this is not possible. Nevertheless the symmetries of the surface BZ follow from the projection of the bulk reflection planes [see panel (d)]. The brown shaded areas in panels (b), (c), and (d) indicate the bulk mirror planes.
 The lower panels show the different surface terminations. The crystal symmetries of the (001), (011), and (111) surfaces are described by the 2D space groups $p4m$, $pmm$, and $p3m1$, respectively.}   \label{structure}
\end{figure*}

\noindent
{\bf Topology of anti-perovskite materials.}
As an example of a TCI with type-II Dirac surface states, we consider
the cubic anti-perovskite materials A$_3$EO with space group $Pm \bar{3} m$ (Table~\ref{mTab1}).
The crystal structure of  A$_3$EO is an inverse perovskite structure, where the oxygen atom O is surrounded octahedrally
by the alkaline earth metal atoms A [see Fig.~\ref{structure}(a)].
We choose Ca$_3$PbO as a generic representative  of this materials class.
The bulk band structure of Ca$_3$PbO displays six Dirac cones, which are gapped by spin-orbit coupling.
While  Ca$_3$PbO is known to be a trivial TR invariant insulator~\cite{kariyadoJPSJ12} (i.e., a trivial class AII insulator~\cite{chiu_RMP_16}), it has recently been argued that reflection symmetries give rise to a nontrivial wavefunction topology with  non-zero mirror Chern numbers~\cite{TCI_Fu_antiperovskites}.

 
 Let us now discuss in detail how the non-trivial topology of  A$_3$EO  arises due to reflection symmetry. First, we observe
that the space group  $Pm \bar{3} m$ possesses    nine different reflection symmetries $R_i$ which transform
${\bf r} = (x,y,z)$ as (see Fig.~\ref{structure})
\begin{subequations}
\begin{align}
R_{x}{\bf r}=&(-x,y,z),& R_{y,\pm z}{\bf r}=&(x,\pm z, \pm y), \\
R_{y}{\bf r}=&(x,-y,z) ,& R_{z,\pm x}{\bf r}=&(\pm z, y, \pm x), \\ 
R_{z}{\bf r}=&(x,y,-z) ,& R_{x,\pm y}{\bf r}=&(\pm y, \pm x, z) .
\end{align}
\end{subequations}
By Fourier transforming into momentum space, we find that there are 12 mirror planes in 
the Brillouin zone (BZ), namely, $k_i=0, \pi$ and $k_i=\pm k_j$ for $i,j=x,\ y,\ z$ and $i\neq j$. 
For each of these reflection planes we can define a mirror Chern number~\cite{Chiu_reflection,Morimoto2013,ChiuSchnyder14}. 
However, due to the cubic rotational symmetries, only 3 out of these 12  mirror invariants are independent. 
Without loss of generality, we choose as an independent set the mirror Chern numbers 
 $C_{x^0}$, $C_{x^\pi}$, and $C_{x,y}$  that are defined
for the reflection planes $k_x=0$, $k_x=\pi$, and $k_x = k_y$, respectively.
Both first-principles calculations and low-energy effective considerations
show that for the cubic anti-perovskites the mirror Chern numbers take the values $C_{x^0}=C_{x,y}=2$ and $C_{x^\pi}=0$ (see Methods and Table~\ref{mTab1}).
Thus, in total there are nine nonzero mirror Chern numbers, i.e.,  $C_{i^0} = C_{i,\pm j} = 2$ for $i, j= x,y,z$ and $i \ne j$.
As shown in Appendix~\ref{low energy}, the low-energy description of A$_3$EO  is given by
six gapped Dirac cones.  Within this low-energy model one finds that
there exists only one bulk gap term $m$ which respects the reflection symmetries
and which gaps out all six Dirac cones.
%
%
%
%
The sign of this gap opening term, $\sgn (m)$, determines the mirror Chern numbers, i.e.,
 \begin{eqnarray}
 &&
 C_{x,y}=  \sgn (m) + b_{x,y},\quad C_{x^0}= 2 \sgn (m)  + b_x,
 \end{eqnarray}
where $b_{x,y}$ and $b_x$ are the mirror Chern numbers of the ``background" bands, i.e.,
those filled bands that are not included in the low-energy description of the bulk Dirac cones.
For the cubic anti-perovskites we find that $b_{x,y}=1$ and $b_x=0$. 
Hence,  $C_{x^0}$ is always non-zero even if the sign of the gap term switches.

\begin{figure}[t]
\includegraphics[clip,width=0.98\columnwidth]{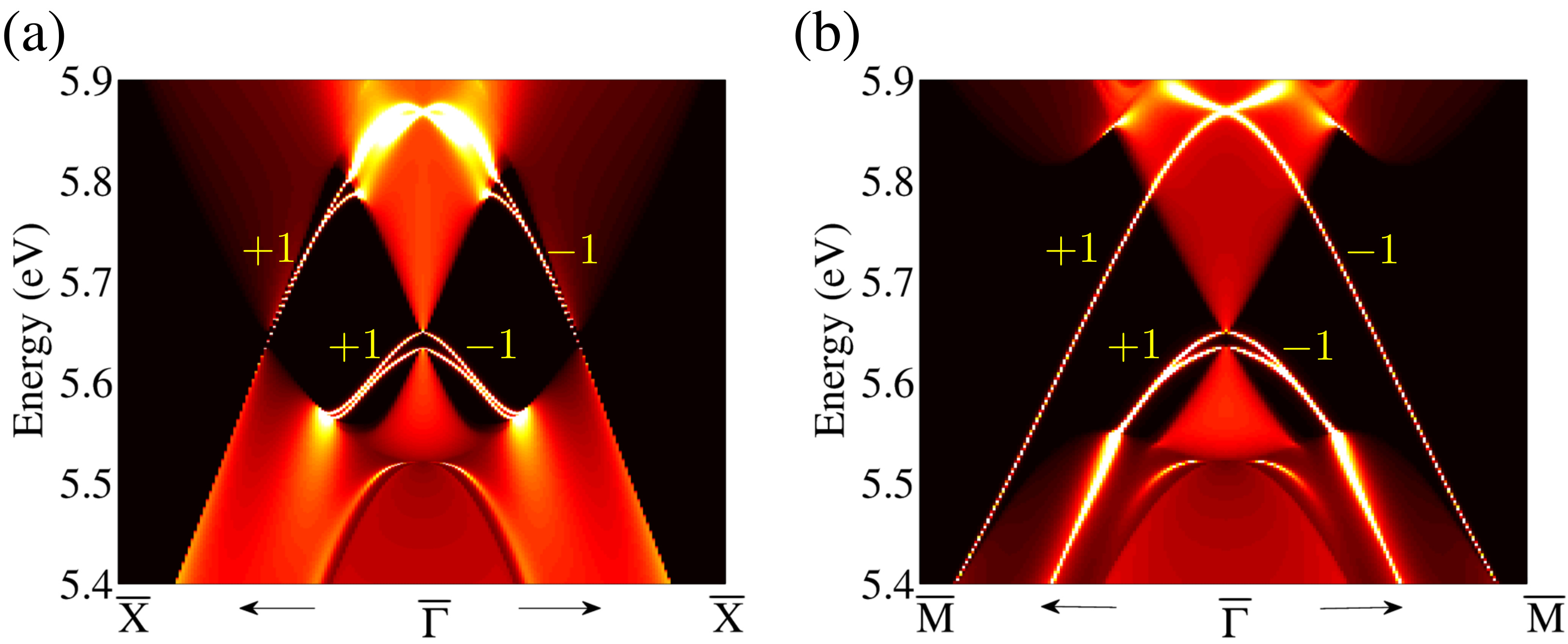}
 \caption{{\bf Density of states at the (001) surface.}
(a),~(b) Surface density of states along the high-symmetry lines $\bar{\Gamma} \to \bar{X}$ and $\bar{\Gamma} \to \bar{M}$ of
the (001) surface BZ, respectively, for the Pb-Ca termination  [lower left panel in Fig.~\ref{mFig1}(b)].
The two high-symmetry directions correspond to the two inequivalent mirror lines of Fig.~\ref{mFig1}(b), i.e.,
$k_y=0$ and $k_x = k_y$.
The spectrum along the mirror line $k_y=0$ [panel (a)] exhibits two right-moving chiral modes with mirror eigenvalue $R_y=+1$ and two left-moving chiral modes with mirror eigenvalue $R_y=-1$. Similarly, the spectrum along the mirror line $k_y=k_x$  [panel (b)] shows two right-moving chiral modes with $R_{x,-y}=+1$ and two left-moving chiral modes with $R_{x,-y}=-1$. 
We observe that the Dirac nodal points (i.e., the bandcrossings) for the upper chiral modes are hidden by the bulk bands.
For the lower chiral modes there is no band crossing at all, since the modes are too far apart.
}   
  \label{001}
\end{figure}

\noindent
{\bf Surface states.} 
By the bulk-boundary correspondence,
a nontrivial value of the mirror Chern numbers $C_{i^0} $ (or $C_{i,\pm j}$)
leads to the appearance of Dirac states on those surfaces that are left invariant 
by the corresponding mirror symmetry $R_{i^0}$ (or $R_{i,\pm j}$). That is,
the value of $C_{i^0} $ (or $C_{i,\pm j}$) indicates
the number of  left- and right-moving chiral modes in the surface BZ.
These chiral surface modes are located within the mirror line $k_{i} =0$ (or $k_i = \pm k_j$)
 that is symmetric under  the reflection operation $R_{i^0}$ (or $R_{i,\pm j}$), see Fig.~\ref{mFig1}.
Importantly,  left- and right-moving  surface chiral modes belong
to opposite eigenspaces of the reflection operators $R_{i^0}$ (or $R_{i,\pm j}$),
and therefore cannot hybridize, see Fig.~\ref{001}. That is, the band crossing between the
 left- and right-moving modes is protected by reflection symmetry. 
Depending on the surface orientation of  A$_3$EO, this band crossing corresponds to the Dirac point of a type-I surface state, the touching
of electron- and hole-pockets of a type-II Dirac state, or is completely hidden in the bulk bands.
In the following, we discuss these three possibilities for the case of Ca$_3$PbO. 


\noindent
{\bf Hidden Dirac nodes on the (001) surface.} 
We start by examining the Dirac states on the (001) surface for the Ca-Pb termination [lower left panel in Fig.~\ref{mFig1}(b)].
Projecting the symmetries of  $Pm\bar{3}m$ along the (001) direction, one finds that the two-dimensional space 
group of the (001) surface is $p4m$~\cite{2D_classification_Liu}. The wallpaper group $p4m$ contains four reflection
symmetries, i.e., $R_x$, $R_y$, $R_{x, + y}$, and $R_{x, - y}$ [Fig.~\ref{mFig1}(b)]. In the surface BZ,
this gives rise to  six mirror lines $k_x=0, k_x=\pi, k_y=0, k_y=\pi, k_x=\pm k_y$ with three independent 
mirror Chern numbers $C_{x^0}$, $C_{x^\pi}$, and $C_{x,y}$. From the above
analysis we find that $C_{x^0}=C_{y^0}=2$ and $C_{x,y} = C_{x,-y} = 2$, which leads to two pairs of
  left- and right-moving chiral modes within the mirror lines $k_x=0$, $k_y=0$ and $k_x=+ k_y$, $k_x=-k_y$, 
  respectively. The left- and right-moving chiral modes belong to 
  reflection eigenspaces with $R=-1$ and $R=+1$, respectively. This is clearly visible in Fig.~\ref{001}, which shows the
  surface density of states along the high-symmetry lines $\bar{\Gamma}  \to \bar{X}$ and 
$\bar{\Gamma}  \to \bar{M}$, which correspond to the mirror lines  $k_x = 0$ and $k_x = k_y$, respectively.
Interestingly, the band crossing formed by the upper left- and right-moving chiral modes is completely
hidden by the bulk bands. The lower chiral modes, on the other hand, do not show a band crossing, 
since they are too far apart.  Similar chiral modes also appear for the Ca-O termination, see Fig.~\ref{mFigS4}.



 %

\begin{figure*}[t]
\includegraphics[clip,width=1.66\columnwidth]{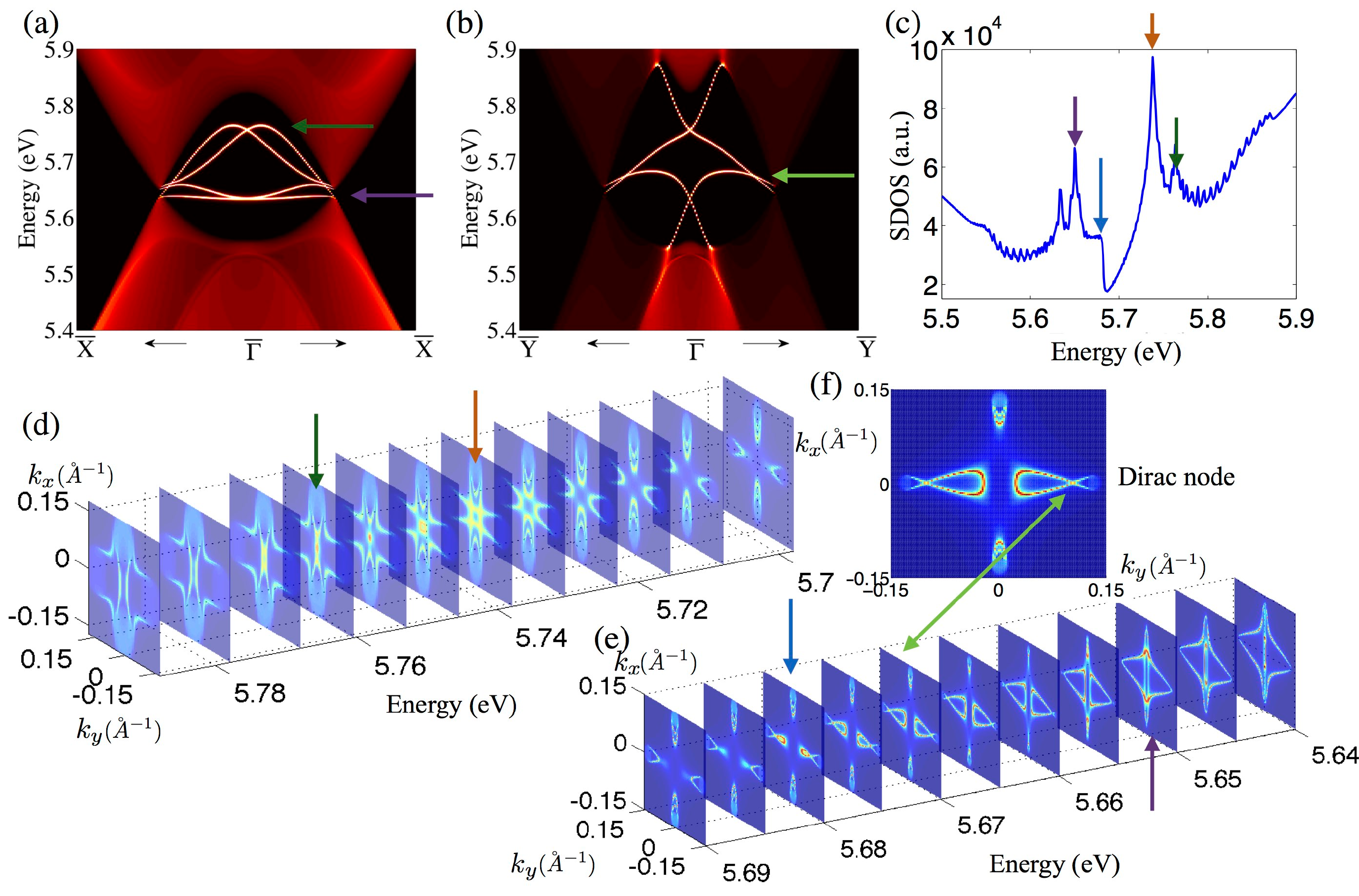}
 \caption{{\bf Density of states at the (011) surface.} 
(a), (b) Surface density of states along the high-symmetry lines $\bar{\Gamma} \to \bar{X}$ and $\bar{\Gamma} \to \bar{Y}$ of the (011) surface BZ, respectively, for the Ca-Pb-O termination [lower left panel in Fig.~\ref{mFig1}(c)]. The two high-symmetry lines represent the intersection of the (011) surface plane with the mirror planes $k_x = 0$ and $k_y = k_z$ of Fig~\ref{mFig1}(c).  The corresponding mirror symmetries protect
the surface band crossings that are visible in panels (a) and (b).
Panel (c) shows the energy resolved surface density of states, which exhibits several van Hove singularities as indicated by the
green, orange, blue, and purple arrows. 
Panels (d) and (e) display the energy- and momentum-resolved surface density of states for different energy ranges.   The van Hove singularities are marked by the arrows. Panel (f) shows the momentum-resolved surface density of states at the energy of
the type-II Dirac point, which is marked by the green arrow.
%
 }   \label{110}
\end{figure*}

\begin{figure*}[t]
\includegraphics[clip,width=1.75\columnwidth]{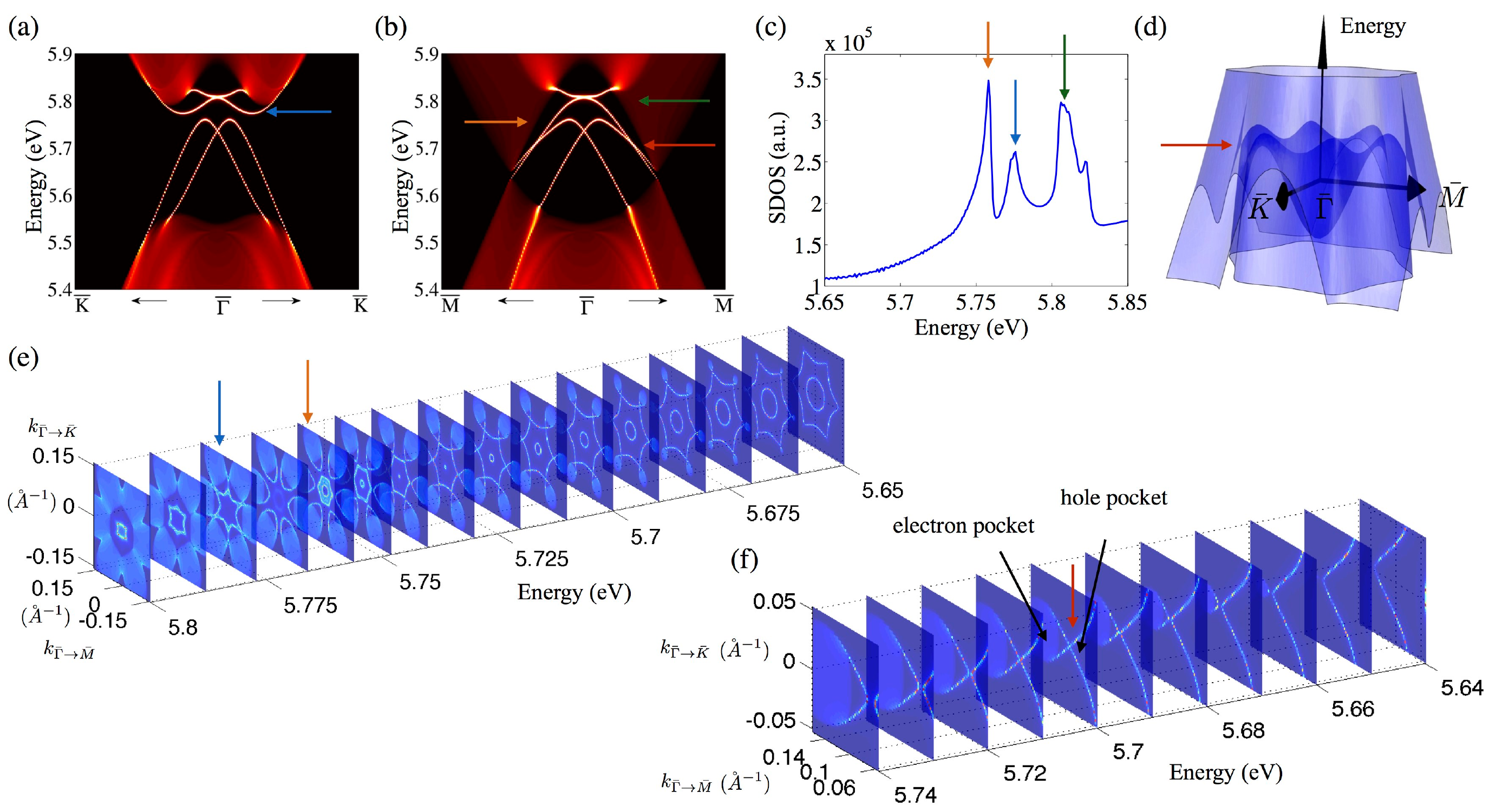}
 \caption{{\bf Density of states at the (111) surface.}
 (a), (b) Surface density of states along the high-symmetry lines $\bar{\Gamma} \to \bar{K}$ 
 and $ \bar{\Gamma} \to \bar{M}$ of the (111) surface BZ, respectively, for the Ca-Pb termination.  
 The $ \bar{\Gamma} \to \bar{M}$ direction corresponds to the mirror lines of the (111) surface. Thus,
 the spectrum along $ \bar{\Gamma} \to \bar{M}$  is gapless and there appear two type-II Dirac states protected
 by reflection symmetry.
The spectrum along the $\bar{\Gamma}-\bar{K}$ line, on the other hand, is gapped, since it is not a mirror line.
(c) Surface density of states which exhibits several van Hove singularities as indicated by the orange, blue, and green arrows. 
The van Hove singularity at $E = 5.76$~eV (orange peak) stems from the back bending of the type-II Dirac state.
 (d) Schematic illustration of the six type-II Dirac states on the (111) surface.
 (e), (f) Energy- and momentum-resolved surface density of states for different energy ranges.
 The van Hove singularities are indicated by the blue and orange arrows. The red
 arrow indicates the type-II Dirac point at  $E=5.70$~eV, where the electron and hole pockets meet. 
}
    \label{111}
\end{figure*}

\noindent
{\bf Type-II Dirac nodes on the (011) surface.} 
Next, we consider the Dirac states on the (011) surface, whose wallpaper group is $pmm$. 
We focus here on the Ca-Pb-O termination; the results for the other termination are shown in Appendix~\ref{appendixD}.
The two-dimensional space group $pmm$ contains  two 
reflection symmetries, $R_x$ and $R_{y,z}$, see Fig.~\ref{mFig1}(c). Correspondingly,
the (011) surface BZ exhibits two mirror lines, namely, $k_x=0$ and $k_{y}=k_z$, with 
the two non-zero mirror Chern numbers $C_{x^0}$ and $C_{yz}$. 
Since $C_{x^0} = C_{yz} = 2$, there appear two pairs of left- and right-moving chiral surface modes within
the mirror lines $k_x=0$ and $k_{y}=k_z$, see Fig.~\ref{110}.
Hybridization between the left- and right-moving chiral modes is prohibited, since they belong to different eigenspaces
of the reflection operators. As indicated by the light green arrows in Figs.~\ref{110}(b), \ref{110}(f), and \ref{110}(e), the two chiral modes at the (011) surface
cross each other at $E=5.67$~eV and form a type-II Dirac point. 
In the close vicinity of this type-II Dirac point the velocity of the two chiral modes has the same sign. 
However, one of the two surface modes
needs to bend over in order to connect bulk valence and conduction bands. This leads
to a maximum and therefore a van Hove singularity in the dispersion of the surface modes. 
The latter reveals itself in the surface density of states as a kink at $E=5.68$ eV, see blue arrows in
Figs.~\ref{110}(c) and~\ref{110}(e). This feature in the surface density of states can 
be used as an experimental fingerprint of the type-II Dirac state.
Another key feature of type-II Dirac points is the touching of electron- and hole-pockets,
see light green arrow at $E=5.67$~eV in Figs.~\ref{110}(e) and~\ref{110}(f).

	
%

	
	
Besides the type-II Dirac nodes at $E=5.67$~eV, there are also two accidental band crossings at the $\bar{\Gamma}$ point of the
surface BZ. These band crossings can be removed by an adiabatic deformation of the surface states. Associated with these accidental Dirac nodes are three van Hove singularities. First, the band crossing at $E=5.75$~eV realizes a van Hoves singularity, which leads to a divergence in the density of states [orange arrow in Figs.~\ref{110}(c) and~\ref{110}(d)]. Second,
the maximum at $E=5.76$~eV in the dispersion of the surface states gives rise to a kink in the surface density of states
[dark green arrow in Figs.~\ref{110}(a) and~\ref{110}(c)]. Third, the flat dispersion of the surface states
near $E=5.65$~eV leads to a peak in the density of states [violet arrow in Figs.~\ref{110}(a), \ref{110}(c), and \ref{110}(e)].



%
%
%
%
%
%
%
%
%
%
%
%
%
%

\noindent
{\bf Type-I and type-II Dirac nodes on the (111) surface.} 
Finally, we examine the Dirac states on the (111) surface for the Ca-Pb termination.
(We note that the surface states of the O termination are expected to be similar to the ones of the Ca-Pb termination, since the oxygen bands are far away in energy from the Fermi energy.) Projecting
the three-dimensional space group $Pm\bar{3}m$ along the (111) direction, we find
that the wallpaper group for the (111) surface is $p3m1$.
The two-dimensional space group $p3m1$ contains three reflection symmetries, i.e., 
 $R_{x,y}$, $R_{x,z}$, and $R_{y,z}$. The corresponding mirror lines in the surface
 BZ are $k_x = k_y$, $k_x = k_z$, and $k_y = k_z$, i.e., the $\bar{M}- \bar{\Gamma}-\bar{M}$  lines.
 For each of the three mirror lines one can define a mirror Chern number $C_{i,j}$,
which are related to each other by the three-fold rotation symmetries of $Pm\bar{3}m$.
As discussed in the Methods, we find that $C_{x,y} = C_{x,z} = C_{y,z} =2$.
By the bulk-boundary correspondence, it follows that there appear two pairs of left- and right-moving
chiral modes within the $\bar{M}- \bar{\Gamma}-\bar{M}$ lines of the surface BZ, see Fig.~\ref{111}(b).
These chiral bands cross each other at $E=5.70$~eV, thereby forming type-II Dirac points [red arrow in Fig.~\ref{111}(b)]. Close to these type-II Dirac points, the velocities of the chiral modes have the same sign. But further away, one of the two modes bends  
 over, such that it connects bulk valence and conduction bands. Hence, this  surface band must exhibit a maximum [orange arrows in Figs.~\ref{111}(b) and~\ref{111}(e)], which leads to a van Hove singularity in the surface density of states at
 $E=5.76$~eV [orange arrow in Fig.~\ref{111}(c)]. Another key feature of this type-II Dirac state is the 
touching of the electron and hole Fermi surfaces. That is, with increasing Fermi energy the open electron and hole-pockets
 approach each other, touch at the type-II Dirac point with $E=5.70$~eV [red arrow in Fig.~\ref{111}(f)], and then 
 separate again.
 

In addition to these type-II Dirac nodes, the (111) surface also exhibits two accidental type-I Dirac nodes at the $\bar{\Gamma}$ point, which can be removed by adiabatic transformations. Connected to these accidental Dirac nodes are two van Hove singularities.
First, the Dirac point at $E=5.81$~eV represents a saddle-point van Hove singularity, which leads to a log divergence
in the surface density of states [green arrows in Figs.~\ref{111}(b) and ~\ref{111}(c)].
Second, the lower bands of the Dirac state at $E=5.81$~eV   bend over, forming a minimum at $E=5.78$~eV. This leads to a kink
in the surface density of states [blue arrow in Figs.~\ref{111}(a) and~\ref{111}(c)].  
 
%

\begin{figure}[!]
\includegraphics[scale=0.72]{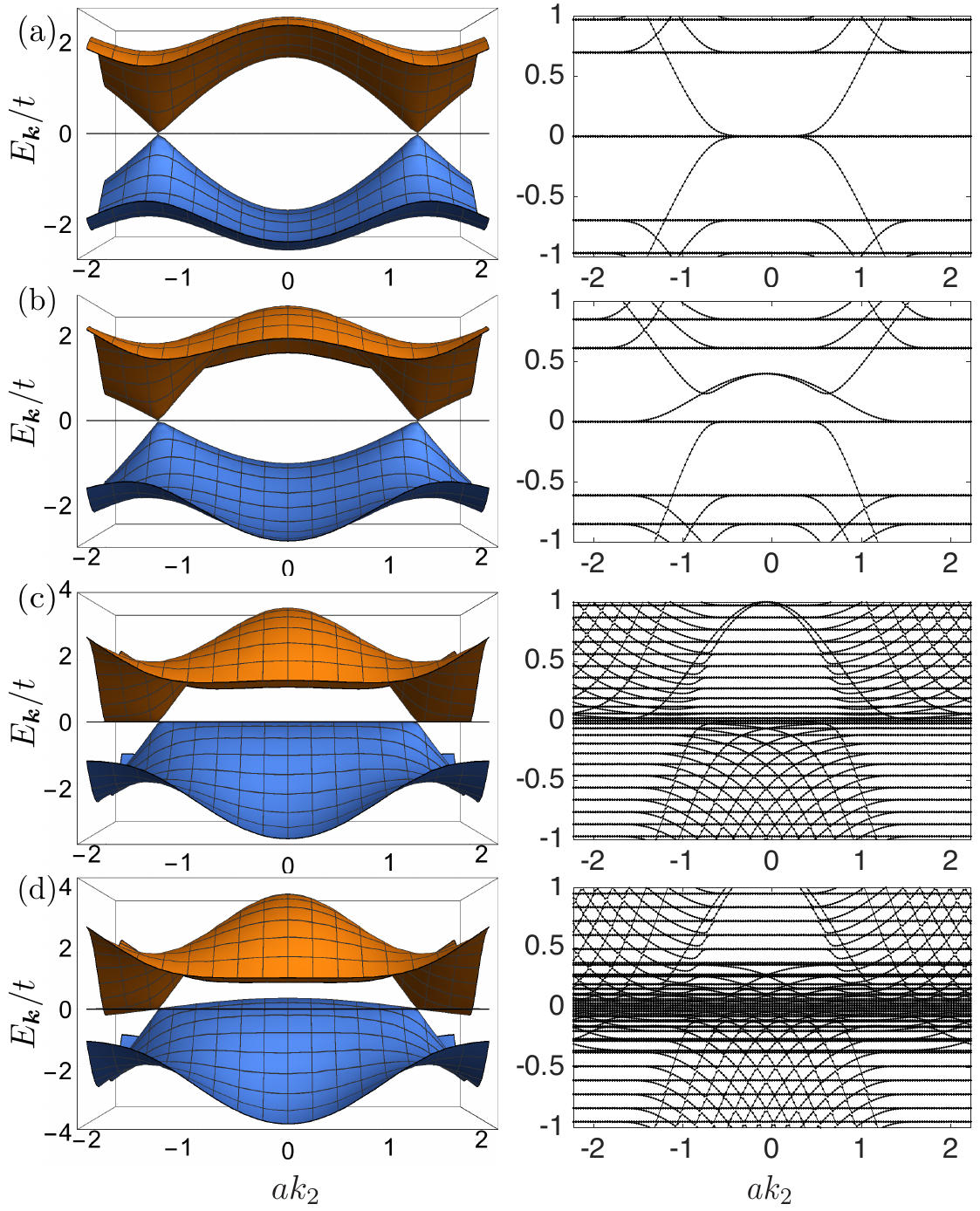}
\caption{\label{Fig:ToyModel}  
\textbf{Landau level spectrum for type-I and type-II Dirac surface states.}
{The left and right columns show the
energy spectrum  and Landau level structure, respectively, for the tight-binding model given by Eq.~\eqref{ham_tilting_main}.
The magnetic field strength is chosen to be $B= 0.01 \phi_0 / a^2$, where $\phi_0 \equiv h/e$ is the flux quantum
and $a^2$ is the are of a plaquette of the square lattice, see Fig.~\ref{Fig:Lattice}.
The tilting of the Dirac cones is controlled by the ratio $t_1 / t$.
The four rows correspond to (a) $t_1/t = 0$ (type-I Dirac cone), (b) $t_1/t = 0.2$ (type-I Dirac cone), (c) $t_1/t = 0.5$ (transition between type-I and -II Dirac cones), and (d) $t_1/t = 0.6$ (type-II Dirac cone), respectively. 
All dispersive curves in the Landau level structure arise from states localized at the edge of the sample, and thus are not of interest here. }
} \label{Landau}
\end{figure}

\noindent
\textbf{Landau level spectrum.} 
A drastic difference between type-I and type-II Dirac surface states arises when a magnetic field is applied. 
%
%
For type-I Dirac cones the energy spectrum of the Landau levels is given by
\bee
E_n \sim \sqrt{n},
\ee 
where $n$ is the Landau level index. Hence, the Landau levels of type-I Dirac cones are in general well separated. 
This is in contrast to type-II Dirac cones.
To illustrate this, let us consider the following tight-binding model on a square lattice~\cite{Kawarabayashi2011}, 
\begin{align} \label{ham_tilting_main}
	H = -\sum_{\average{ij}} t_{ij} b_j^\dagger a_{i} + t_1 \sum_{\average{ij}} \left(a_j^\dagger a_i +b_j^\dagger b_i \right),  
\end{align}
where $a_i$ and $b_i$ denote the electron annihilation operators on the sublattice A and B, respectively. 
$t_{ij}$ represent the nearest neighbor hopping, while $t_1$ is the next nearest neighbor hopping integral
(for more details see Appendix~\ref{appendixC}).
Hamiltonian~\eqref{ham_tilting_main} describes two Dirac cones, whose tilting is
controlled by the ratio $t_1 / t$, with $t = \left| t_{ij} \right|$.
For $t_1 / t = 0$ there is no tilting [Fig.~\ref{Fig:ToyModel}(a)]  and 
for $t_1 / t = 0.2$ there is a small tilting [Fig.~\ref{Fig:ToyModel}(b)]. 
In both cases there exist well separated Landau levels.
(Note that the dispersive curves in the Landau level structure are due to edge states
and therefore should be ignored in the following discussion.)
At $t_1 / t = 1/2$ there is a transition from type-I Dirac states to type-II Dirac states. At this transition point the 
spectrum becomes nondispersive along the $k_2$ direction and the Landau levels get very dense around
$E=0$ (i.e., around the energy of the Dirac point) [Fig.~\ref{Fig:ToyModel}(c)].
Finally, for $t_1/t > 1/2$, there appear type-II Dirac cones with open electron and hole pockets.   
As shown in Fig.~\ref{Fig:ToyModel}(d), for type-II Dirac cones the separation between the Landau levels near $E=0$ is close to zero,
leading to a sizable region of very dense Landau levels  [cf. Fig.~\ref{Landau}(d)].
This region of dense Landau levels arises because the open electron and hole Fermi surfaces enclose a very large momentum-space area,
which is much larger than the one enclosed by type-I Dirac states (for a detailed explanation, see Appendix~\ref{appendixC}). 
Moreover, we find that for type-II Dirac states the zeroth Landau level $E_0$ of the type-I Dirac node [see Figs.~\ref{Fig:ToyModel}(a) and~\ref{Fig:ToyModel}(b)] 
becomes unpinned and moves away from $E=0$.


%

%

%
%
%
%
%
%
%
%
%
%


\vspace{0.5cm}

\noindent
\textbf{\large Discussion}\\
Using general symmetry arguments, we have shown that the surface of crystalline topological insulators can 
host type-II Dirac surface states, which are characterized by open electron and hole Fermi surfaces that touch each other.
 This is in contrast to regular strong topological insulators, where the Dirac surface states, due to time-reversal symmetry,
are always of type-I, which exhibit a closed small Fermi surface.
By means of ab-initio DFT calculations, we have demonstrated that type-II Dirac states appear at the surfaces of the antiperovskite materials A$_3$EO.
As a representative example, we have considered Ca$_3$PbO and determined the surface spectra for the (001), (011), and (111) surfaces. 
The type-II Dirac nodes appear 
on the (011) surface with Ca and Ca-Pb-O terminations, and on the (111) surface with  Ca-Pb termination (see Figs.~\ref{110} and~\ref{111}).
All these band crossings are protected by reflection symmetry. That is, the left moving and right moving chiral modes that cross
each other belong to different eigenspaces of the reflection operator $R$.
We have shown that type-II Dirac surface states possess van Hove singularities, since one of the two chiral modes needs to bend over in order to connect valence with conduction bands. These van Hove singularities lead to divergences and kinks in the surface density of states, which can serve as unique fingerprints of the type-II Dirac states.
Another distinguishing feature of type-II Dirac states is their Landau level spectrum. As opposed to type-I Dirac states, where the Landau levels are well separated, for type-II Dirac states there exists a very large density of Landau levels near the band-crossing energy (see Fig.~\ref{Landau}).
It will be interesting to compare these theoretical findings with quantum oscillations~\cite{rost_to_be_published}, angle-resolved photoemission, and scanning tunneling experiments.

Using a low-energy theory and a DFT-derived tight-binding model, we have determined the mirror Chern numbers for the cubic antiperovskites.  We have shown that the mirror Chern numbers for the $k_i=0$ and $k_i=\pm k_j$ (for $i,j=x,\ y,\ z$ and $i\neq j$) mirror planes are equal to two,
indicating that there appear two left- and right-moving modes on surfaces that are invariant under the mirror symmetries.
Depending on the surface orientation and termination these left- and right-moving modes form  type-I or type-II Dirac nodes, or do not
cross at all.
We remark that while the mirror Chern numbers determine the number of left- and right-moving chiral modes 
that connect valence and conduction bands, they do not give any information about the number of Dirac
band crossings in the surface spectrum. This is because, (i) the Dirac points might be hidden in the bulk,
(ii) the left- and right-moving modes might be too far apart to form a crossing, or (iii) there might be
accidental band crossings.

	



\vspace{0.5cm}

\noindent
\textbf{\large Methods}\\

\noindent
\textbf{DFT calculations and tight-binding model.} 
The electronic band structure of the cubic antiperovskites A$_3$EO is determined by performing first-principles calculations with the Vienna ab initio package \cite{Kresse1996,Kresse1996-2} using the projector augmented wave (PAW) method \cite{Bloechl1994,Kresse1999}.
As an input for the DFT calculation we used the experimental crystal structure of Ref.~\cite{Nuss:dk5032}. 
The lattice constant for Ca$_3$PbO is 4.847 \r{A}.
For the exchange-correlation functional 
we chose the generalized-gradient approximation of Perdew-Burke-Ernzerhof type~\cite{pbe_PRL_96}.
The plane wave basis is truncated with an energy cut-off of 400 eV.
For the bulk calculation a $12\times12\times12$ k-mesh is used. Spin-orbit coupling effects are also taken into account. 

The DFT calculations show that near the Fermi energy $E_F$ the valence bands mostly originate from Pb-$p$ orbitals ($p_x$, $p_y$, and $p_z$), while the orbital character of the conduction bands near $E_F$ is Ca-$d_{x^2-y^2}$, Ca-$d_{xz}$, and Ca-$d_{yz}$
(from three different Ca atoms).
Guided by these findings, we use these 12 orbitals (24 including spin) as a basis set to derive a low-energy tight-binding model.
We determine the hopping parameter values for this tight-binding model
 from a maximally localized Wannier function (MLWF) method~\cite{Mostofi2008,marzariRMP12}.
 With this model, we compute the momentum-resovled surface density of states by means of
 an iterative Green's function method~\cite{sancho_iterative_Green}. The results of these calculations are shown in Figs.~\ref{001}, \ref{110}, 
 and \ref{111}.
 To determine the topological characteristics of Ca$_3$PbO we have also used a simplified nine-band (18 bands including spin) tight-binding model, see Appendix~\ref{TB_model} for details.

\vspace{0.2cm}

\noindent
\textbf{Topological invariants.} 
The type-I and type-II Dirac surface states of the cubic antiperovskites are protected by a mirror Chern number. 
The mirror Chern number is defined as a two-dimensional integral over the reflection plane of the occupied wave functions 
with mirror eigenvalue $R=+1$ (or $R=-1$)~\cite{Teo:2008fk,Hsieh:2012fk}.
Note that since the Hamiltonian $H$ commutes with the reflection operator, the eigenfunctions of $H$ can be assigned a definite mirror eigenvalue.
Without loss of generality, one usually assumes that the mirror eigenvalues are $\pm 1$, since $R^2=1$ after a suitable 
$U(1)$ gauge transformation~\cite{Chiu_reflection}. 
The value of the mirror Chern number  corresponds to the number of  left- and right-moving chiral surface modes. 
These chiral surface modes exist within the mirror line of the surface BZ, i.e., within the line that is obtained by projecting
the bulk mirror plane onto the surface BZ.

We have numerically computed the mirror Chern number using two different methods: (i) using the simplified tight binding model 
of Appendix~\ref{TB_model}
and (ii) using the real space 
wavefunctions of the DFT-derived 12-band tight-binding model.
For method (i) the reflection operator $R$ can be written explicitly in momentum space. The momentum space Hamiltonian can then be block diagonalized with respect to $R$ and the eigenfunctions can be obtained for each block separately
(see Appendix~\ref{simple model} for details). For method (ii)
the real-space wavefunctions of the 12-band tight-binding model are projected onto the mirror eigenspaces $(\bI\pm R)/2$. 
This is done by identifying mirror-reflected orbitals with proper sign changes.
Using these projected wavefunctions a Fourier transform is performed along the two surface momenta to obtain the surface spectrum
for a given reflection eigenspace. The Chern number can then be inferred 
from the number of chiral surface modes in the surface spectrum.
Both methods (i) and (ii) agree with each other. 


 

\vspace{0.5 cm}
 
\noindent
\textbf{Acknowledgements} \\
 We gratefully acknowledge many useful discussions with M.~Y.~Chou, M.~Franz, M.~Hirschmann, H.~Nakamura, J.~Nuss, A.~Rost, and  H.~Takagi. 
C.K.C.\ would like to thank the Max-Planck-Institut FKF Stuttgart for its hospitality and acknowledge the support of the Max-Planck-UBC Centre for Quantum Materials and Microsoft. X.L. and C.K.C. are supported by LPS-MPO-CMTC.
Y.H.C. is supported by a Thematic project at Academia Sinica.\\

\vspace{0.5 cm}
 
\noindent
\textbf{Competing financial interests} \\
 The authors declare that they have no
competing financial interests.

\vspace{0.5 cm}
 
\noindent
\textbf{Author contributions} \\
Y.H.C. and Y.N. performed the ab-initio first-principles calculations.
 All authors contributed to the discussion and interpretation of the results and to the writing of the paper.
 
\vspace{0.5 cm}
 
\noindent
\textbf{Data availability} \\
All relevant numerical data are available from the authors upon request.



\appendix

\setcounter{figure}{0}
\makeatletter 
\renewcommand{\thefigure}{S\@arabic\c@figure} 

\makeatother

\bibliography{TOPO3_v13,DFT}

 \clearpage
\newpage

\begin{center}
\textbf{
\large{Supplementary Information for}}
\vspace{0.4cm} 

\textbf{
\large{
``Type-II Dirac surface states in topological crystalline insulators" } 
}
\end{center}

\vspace{0.1cm}

\begin{center}
\textbf{Authors:} 
Ching-Kai Chiu,
Y.-H. Chan,
Xiao Li,
Y. Nohara,
 and
 A. P. Schnyder
\end{center}

\vspace{0.5cm}

In this supplementary information we present the details of the simplified nine-band tight-binding model, a low-energy effective theory of the
cubic antiperovskites, their Landau level structure, and the surface state spectra for some additional surface terminations.

\section{Simplified tight-binding model of Ca$_3$PbO} \label{simple model} \label{TB_model}

To construct a simplified tight-binding model we follow along the lines of the work by Kariyado and Ogata~\cite{kariyadoJPSJ12}.
In Ref.~\cite{kariyadoJPSJ12} a six-band model with the orbitals
 \begin{align*}
\mathrm{Pb}_{p_x},\mathrm{Pb}_{p_y}, \mathrm{Pb}_{p_z},
\mathrm{Ca}^1_{d_{y^2-z^2}},\mathrm{Ca}^2_{d_{z^2-x^2}}, \textrm{and} \; \mathrm{Ca}^3_{d_{x^2-y^2} } 
\end{align*}
was constructed. This six-band model exhibits six  \emph{gapless} Dirac  nodes
along the $\Gamma-X$ direction, but does not contain a Dirac mass gap, which is present in the DFT calculations. 
To open up a gap one needs to include in addition the Ca$^1$-$d_{yz}$, Ca$^2$-$d_{zx}$, and Ca$^3$-$d_{xy}$ orbitals.
As we show below, the spin-orbit coupling between these orbitals and the Ca$^1$-$d_{y^2-z^2}$,
Ca$^2$-$d_{z^2-x^2}$, and Ca$^3$-$d_{x^2-y^2}$ orbitals 
represents a mass term, that opens up a gap at the six Dirac cones. 
We use this nine-band model to analyze the topological properties 
of Ca$_3$PbO (and other cubic antiperovskites) and  to compute the mirror Chern numbers.


Thus, in the absence of spin-orbit coupling our tight-binding Hamiltonian is written as
$\mathcal{H} = \sum_{\bf k}   \psi^{\dag}_{\bf k}  H_{k/2}  ( {\bf k} ) \psi_{\bf k} $
with the ninne-component spinor
\begin{align*}
\psi_{\bf k}=
(&\mathrm{Pb}_{p_x},\quad\quad\mathrm{Pb}_{p_y}, \quad\quad\mathrm{Pb}_{p_z}, \\
&\mathrm{Ca}^1_{d_{y^2-z^2}},\quad \mathrm{Ca}^2_{d_{z^2-x^2}}, \quad\mathrm{Ca}^3_{d_{x^2-y^2}}, \\
&\mathrm{Ca}^1_{d_{yz}},\quad\quad\mathrm{Ca}^2_{d_{zx}},\quad\quad\mathrm{Ca}^3_{d_{xy} } )^{\textrm{T}}
\end{align*}
and the $9 \times 9$ matrix $H_{k/2}$, which can be expressed in block form as
\begin{eqnarray} \label{ham_wo_soc}
H_{k/2} ( {\bf k} )
=
\begin{pmatrix}
H_p & V_{dp}^u & V_{dp}^l \cr
{V_{dp}^u}^{\dag} & H_d^u & 0 \cr
{V_{dp}^l}^\dag & 0 & H_d^l \cr
\end{pmatrix} .
\end{eqnarray}
The blocks of $H_{k/2}  ( {\bf k} )$ are given by
\begin{eqnarray}
H_p
=
\begin{pmatrix}
e_p - 2 t_{pp} c_{2x}  & 0 & 0   \cr
0 & e_p - 2 t_{pp} c_{2y}  & 0  \cr
0 & 0 & e_p - 2 t_{pp} c_{2z}  \cr
\end{pmatrix} 
\end{eqnarray}
and
\begin{eqnarray}
H_d^u
=
\begin{pmatrix}
e_d & - 4 t_{dd} c_x c_y & - 4 t_{dd} c_z c_x \cr
- 4 t_{dd} c_x c_y &  e_d & - 4 t_{dd} c_y c_z \cr
- 4 t_{dd} c_z c_x & - 4 t_{dd} c_y c_z & e_d  \cr
\end{pmatrix}  \nonumber
\end{eqnarray}
and $H_d^l=e_d\bI_3$,  
with $\bI_3$ the 3$\times$3 identity matrix.
The coupling terms between $p$ and $d$ orbitals are
\bee
V_{dp}^u=4 i t_{pd}
\begin{pmatrix}
0 & c_z s_x & - c_y s_x  \cr
- c_z s_y & 0 & c_x s_y \cr
c_y s_z & - c_x s_z& 0 \cr 
\end{pmatrix}, \quad  V_{dp}^l=4 i t_{pd}
\begin{pmatrix}
 0 & c_x s_z & c_x s_y \cr
 c_y s_z & 0 & c_y s_x \cr
 c_z s_y & c_z s_x & 0 \cr 
\end{pmatrix},
\ee
where we have used the short-hand notation
\begin{eqnarray}
c_i = \cos \frac{ k_i }{2}, 
\quad
s_i = \sin \frac{ k_i }{2} ,
\quad
c_{2i} 
= \cos k_i . \nonumber
\end{eqnarray}
In order to simplify matters, we have neglected in the above expressions further neighbor hopping terms that were included in the work
by Kariyado and Ogata~\cite{kariyadoJPSJ12}. We have checked that these simplifications do not alter the topological properties. 


Let us now add spin-orbit coupling terms to the Hamiltonian~\eqref{ham_wo_soc}.
The on-site spin-orbit coupling for the Pb-$p$ orbitals is given by
$\sum_{\bf k} \psi^{\dag}_p ({\bf k} )  H^p_{\mathrm{SO}} ( {\bf k} )  \psi^{\ }_p ({\bf k} ) $ with
the spinor 
\begin{eqnarray}
\psi^{\ }_p ({\bf k} ) = ( \mathrm{Pb}^{\uparrow}_{p_x} ,  \mathrm{Pb}^{\uparrow}_{p_y},  \mathrm{Pb}^{\uparrow}_{p_z}  ,  \mathrm{Pb}^{\downarrow}_{p_x} ,  \mathrm{Pb}^{\downarrow}_{p_y},  \mathrm{Pb}^{\downarrow}_{p_z}   ) \nonumber
\end{eqnarray}
and
\begin{eqnarray}
H^p_{\mathrm{SO}} ( {\bf k} ) 
=
\frac{\lambda_p}{2} 
\begin{pmatrix}
0 & - i & 0 & 0 & 0 & 1 \cr
i & 0 & 0 & 0 &0 & - i \cr
0 & 0 & 0 & - 1 & \; i \; & 0 \cr
0 & 0 & -1 & 0 & \; i \;  & 0 \cr
0 & 0 & - i & - i & 0 & 0 \cr
1 & i & 0 & 0 &0 & 0 \cr
\end{pmatrix} . \nonumber
\end{eqnarray}
The on-site spin-orbit coupling for the $d$ orbitals reads $\sum_{\bf k} \psi^{\dag}_d ({\bf k} )  H^d_{\mathrm{SO}} ( {\bf k} )  \psi^{\ }_d ({\bf k} ) $ with
the spinor 
\begin{eqnarray} 
&& \psi_d ( {\bf k} ) =
\nonumber\\
&& (  \mathrm{Ca}^{1, \uparrow} _{d_{y^2-z^2}} ,  
 \mathrm{Ca}^{2, \uparrow}_{d_{z^2-x^2}},
 \mathrm{Ca}^{3, \uparrow}_{d_{x^2-y^2}} ,
  \mathrm{Ca}^{1, \downarrow} _{d_{y^2-z^2}} ,  
 \mathrm{Ca}^{2, \downarrow}_{d_{z^2-x^2}},
 \mathrm{Ca}^{3, \downarrow}_{d_{x^2-y^2}} ,
\nonumber\\
&&
  \mathrm{Ca}^{1, \uparrow}_{d_{yz}},
  \mathrm{Ca}^{2, \uparrow}_{d_{zx}},
 \mathrm{Ca}^{3, \uparrow}_{d_{xy} } ,
  \mathrm{Ca}^{1, \downarrow}_{d_{yz}},
  \mathrm{Ca}^{2, \downarrow}_{d_{zx}},
 \mathrm{Ca}^{3, \downarrow}_{d_{xy} } ,
 )^{T}, 
 \nonumber
\end{eqnarray}
and
\begin{align}
 H^{d}_{\mathrm{SO}} ( {\bf k} ) 
 =&\lambda_d  \tau_y \otimes \Big \{
 \sigma_x \otimes 
\begin{pmatrix}
  1 & 0 & 0\cr
  0 & 0 & 0 \cr
  0 & 0 & 0 \cr
 \end{pmatrix} \nonumber \\
&+  \sigma_y \otimes 
\begin{pmatrix}
  0 & 0 & 0\cr
  0 & 1 & 0 \cr
  0 & 0 & 0 \cr
   \end{pmatrix} 
  + \sigma_z \otimes 
\begin{pmatrix}
  0 & 0 & 0\cr
  0 & 0 & 0 \cr
  0 & 0 & 1 \cr
 \end{pmatrix}
\Big \},
 \end{align}
where $\tau_\beta$  and $\sigma_\alpha$ represent $d$-orbital ($x_i^2-x^2_j$ and $x_ix_j$) and spin (up and down) 
degree of freedom, respectively. 
As it turns out $H^{d}_{\mathrm{SO}} ( {\bf k} )$ gaps out the bulk Dirac cones.
 
Adding these spin-orbit coupling terms to Eq.~\eqref{ham_wo_soc}, we obtain the full Hamiltonian
\begin{eqnarray} \label{ham_with_SOC}
H_{\mathrm{tot}}^{k/2}  ( {\bf k} ) 
=
\begin{pmatrix}
H_p^{\textrm{tot}}( {\bf k} ) & V_{\textrm{tot}} ( {\bf k} ) \cr
V^{\dag}_{\textrm{tot}} ( {\bf k}Ê)  &  H_d^{\textrm{tot}}  ( {\bf k} ) \cr
\end{pmatrix} , 
\end{eqnarray}
with 
\begin{align}
H_p^{\textrm{tot}} ( {\bf k} ) 
=&
\begin{pmatrix}
H_p & 0 \cr
0 & H_p \cr
\end{pmatrix}
+
 H^p_{\mathrm{SO}} ( {\bf k} ) , \nonumber \\
 H_d^{\textrm{tot}}  ( {\bf k} ) 
=&
\begin{pmatrix}
\sigma_0\otimes H_d^u & 0 \cr
0 & \sigma_0 \otimes H_d^l \cr
\end{pmatrix}
 +
 H^{d}_{\mathrm{SO}} ( {\bf k} ),  \nonumber
\end{align}
and 
\begin{eqnarray}
V_{\textrm{tot}} ( {\bf k} )
=
\begin{pmatrix}
\sigma_0\otimes V^u_{dp} &   \sigma_0\otimes V^l_{dp}   \cr
\end{pmatrix}. \nonumber
\end{eqnarray} 
Note that the outermost grading of $H_p^{\textrm{tot}}$ and $\sigma_0$ in the above expressions corresponds to the spin grading.
The parameters of the above tight-binding model can be determined by fitting to the DFT results.
We have used the following values (in units of eV) 
\begin{eqnarray}
&&
e_p = 0.0, 
\quad
e_d = 2. 0,
\quad
t_{pp} = -0.4,
\nonumber\\
&&
t_{dd} = -0.4,
\quad
t_{pd} = -0.4,
\quad
\lambda_p =0.75,
\quad
\lambda_d = 0.1 .
\nonumber
\end{eqnarray}
We have checked that the nine-band model Hamiltonian~\eqref{ham_with_SOC} exhibits qualitatively
the same surface states as the DFT-derived twelve-band model (cf.~Fig.~\ref{001}). 

Let us now compute the mirror Chern numbers for this model. To this end, we first need to determine the reflection operators.
To remove fractional momenta in the symmetry operators, we first perform a unitary transformation on the 
Hamiltonian~\eqref{ham_with_SOC}, i.e.,
\bee \label{HamTot_after_U}
H^{k} ( {\bf k} )=U^\dagger H^{k/2}_{\textrm{tot}} (\bk) U,
\ee
where $U=\rm{diag}(U^p,U^{d^u},U^{d^l})$, with
\begin{eqnarray}
U_p
&=&
e^{-i(k_x+k_y+k_z)/2}\bI_6, 
\\
U^u_d
&=&
U^l_d=\sigma_0 \otimes 
\bma
e^{-ik_x/2} & 0 & 0 \\
0 & e^{-ik_y/2} & 0 \\
0 & 0 & e^{-ik_z/2} \\
\ema .
\end{eqnarray}
The reflection operator $R_x$ for the mirror symmetry $k_x \to -k_x$  is given by 
\bee
U_{R_x}(\bk)=
\bma 
R_x^p & 0 & 0 \\
0 & R_{x}^{d^u} & 0 \\
0  & 0 & R_x^{d^l} \\
\ema,
\ee
where 
\begin{align}
R_x^p=&\sigma_x \otimes 
\bma 
-1 & 0 & 0 \\
0 & 1 & 0 \\
0 & 0 & 1\\
\ema,
\nonumber\\
R_x^{d^u}=&\sigma_x \otimes 
\bma 
1 & 0 & 0 \\
0 & e^{ik_x} & 0 \\
0 & 0 & e^{ik_x}\\
\ema,  \nonumber \\
 R_x^{d^l}=&\sigma_x \otimes 
\bma 
1 & 0 & 0 \\
0 & -e^{ik_x} & 0 \\
0 & 0 & -e^{ik_x}\\
\ema.
\end{align}
The expression of the reflection operator $R_{x,-y}$ for the mirror symmetry
$(k_x, k_y) \to ( - k_y, - k_x)$ reads
\bee
U_{R_{x,-y}}(\bk)=
\bma 
R_{x,-y}^p & 0 & 0 \\
0 & R_{x,-y}^{d^u} & 0 \\
0  & 0 & R_{x,-y}^{d^l} \\
\ema,
\ee
where
\begin{align}
R_{x,-y}^p=&-\frac{\sigma_x +\sigma_y}{\sqrt{2}} \otimes 
\bma 
0 & 1 & 0 \\
1 & 0 & 0 \\
0 & 0 & -1\\
\ema,\nonumber\\ 
R_{x,-y}^{d^u}=& -\frac{\sigma_x +\sigma_y}{\sqrt{2}} \otimes 
\bma 
0 & e^{ik_x} & 0 \\
e^{ik_y} & 0  & 0 \\
0 & 0 & e^{i(k_x+k_y)}\\
\ema,  \nonumber \\
 R_{x,-y}^{d^l}=&-\frac{\sigma_x +\sigma_y}{\sqrt{2}}\otimes 
\bma 
0 & e^{ik_x} & 0 \\
e^{ik_y} & 0  & 0 \\
0 & 0 & -e^{i(k_x+k_y)}\\
\ema.
\end{align}
These two reflection symmetries act on the Hamiltonian~\eqref{HamTot_after_U} as
\begin{align}
H^k(\bk)=& U_{R_x}^\dagger H^k(-k_x,k_y,k_z) U_{R_x}, \\ 
H^k(\bk)=& U_{R_{x,-y}}^\dagger H^k(-k_y,-k_x,k_z) U_{R_{x,-y}} .
\end{align}
Due to these two mirror symmetries, the bulk wave functions of $H^k ( {\bf k})$ in the mirror planes $k_x=0$ and $k_x =- k_y$, respectively,
can be labelled by the mirror eigenvalues $\pm 1$. Within these two-dimensional mirror planes in momentum space, 
one can compute the Chern number $n_{\pm 1}$ of the occupied bands for each mirror eigenvalue separately. 
The mirror Chern number is then given by $C = ( n_{+1} - n_{-1} )/2$.
Using this approach we have computed the mirror Chern numbers for $H^k ( {\bf k})$. We find that they are consistent
with the number of chiral surface modes as computed from the DFT-derived twelve-band tight-binding model. 

\section{Low-energy effective theory} \label{low energy}

In this section we study the topology of the cubic antiperovskites A$_3$EO using a low-energy effective theory. 
As discussed in the main text, 
the surface states of A$_3$EO
are protected by the nine reflection symmetries
\begin{subequations} \label{mirror_symmetris_AppB}
\begin{align}
R_{k_x}{\bf k}=&(-k_x,k_y,k_z),& R_{k_y,\pm k_z}{\bf k}=&(k_x,\pm k_z, \pm k_y), \\
R_{k_y}{\bf k}=&(k_x,-k_y,k_z),& R_{k_z,\pm k_x}{\bf k}=&(\pm k_z, k_y, \pm k_x), \\ 
R_{k_z}{\bf k}=&(k_x,k_y,-k_z),& R_{k_x,\pm k_y}{\bf k}=&(\pm k_y, \pm k_x, k_z),
\end{align}
\end{subequations}
and the bulk band structure of A$_3$EO exhibits six Dirac cones, which are gapped out
by spin-orbit coupling. These six Dirac cones are located on the 
$\Gamma - X$ high-symmetry lines of the bulk Brillouin zone,
i.e., at 
\begin{equation}
{\bf k}=(\pm\Delta,0,0),\ (0,\pm\Delta,0),\ (0,0\pm\Delta).
\end{equation}
In the absence of spin-orbit coupling, the low-energy physics near these six Dirac cones 
is described by the Hamiltonian~\cite{kariyadoJPSJ12}
\begin{widetext}
\bee \label{low_energy_hamiltonian}
H({\bf k})=
\begin{dcases} 
(k_x\pm \Delta )\six\otimes \siz + k_y \siy \otimes  \siz + k_z \siz \otimes \siz,  & \text{near } {\bf k}=(\mp \Delta, 0 , 0 )\\
k_x\six\otimes \siz + (k_y\pm \Delta ) \siy \otimes  \siz + k_z \siz \otimes \siz,  & \text{near } {\bf k}=(0,\mp \Delta , 0 ) \\
k_x\six\otimes \siz +  k_y \siy \otimes  \siz + (k_z\pm \Delta ) \siz \otimes \siz, & \text{near } {\bf k}=(0,0,\mp \Delta ) 
\end{dcases} .
\ee
\end{widetext}
We observe that
Eq.~\eqref{low_energy_hamiltonian} is invariant under the nine mirror symmetries~\eqref{mirror_symmetris_AppB}.
That is, the Hamiltonian $H({\bf k})$  obeys 
\bee
U_\#^{-1}H({\bf k}) U_\#= H(R_\#{\bf k})
\ee
with the symmetry operators $U_{k_i}=\sigma_i\otimes \six$ and $U_{k_i,\pm k_j}=\frac{\sigma_i\mp\sigma_j}{\sqrt{2}}\otimes \six$.
Moreover,  we note that the six gapless Dirac cones of Eq.~\eqref{low_energy_hamiltonian} are located within the mirror planes $k_i = 0$ ($i=x,y,z$).
Hence, in order to compute the mirror Chern number for these mirror planes, the Dirac nodes need to be gapped out, which occurs due to 
spin-orbit coupling. Within the low-energy model~\eqref{low_energy_hamiltonian}, we find that there exists only one 
symmetry-preserving gap opening term, namely $m\bI \otimes \six$, with $m$ a constant that is independent of ${\bf k}$.
The mass term  $m\bI \otimes \six$ gaps out all six Dirac nodes. As we will see, this turns the system into 
a non-trivial topological crystalline insulator. To show this we need to determine the mirror Chern numbers. 

%

Let us first consider the mirror Chern number in the $k_z=0$ reflection plane. 
The eigenspace of $U_{k_z}$ with mirror eigenvalue $+1$ is spanned by
$\ket{\psi_1}=(1,1,0,0)/\sqrt{2}$ and $\ket{\psi_2}=(0,0,1,-1)/\sqrt{2}$.
Projecting the low-energy Hamiltonian~\eqref{low_energy_hamiltonian} within the reflection plane
$k_z=0$ onto this eigenspace gives 
\bee
h({\bf k})_{k_z}=
\begin{dcases} 
(k_x\pm \Delta )\six + k_y \siy  + m \siz,  & \text{near } (\mp \Delta, 0 , 0 ),\\
k_x\six+ (k_y\pm \Delta ) \siy  +m \siz,  & \text{near } (0,\mp \Delta , 0 ).
\end{dcases}
\ee
We observe that the four Dirac cones, which are located within the mirror plane $k_z =0$ are gapped out by the same
mass term $m \sigma_z$. Since all four Dirac cones 
have the same orientation, a sign change in $m$ leads to 
a Chern number change by $+1$ for all of the four Dirac cones (or $-1$ for all the four Dirac cones).
Hence, the total mirror Chern number $C_{z^0}$ changes
by four, when $m \to - m$. 

Second, we consider the mirror Chern number in the $k_x = -k_y$ reflection plane. The eigenspace
of $U_{k_x, -k_y}$ with mirror eigenvalue $+1$ is spanned by the vectors 
$\ket{\phi_1}=(e^{-i\pi/8},0,0,e^{i\pi/8})/\sqrt{2}$ and $\ket{\phi_2}=(0,e^{-i\pi/8},e^{i\pi/8},0)/\sqrt{2}$.
Projecting Hamiltonian~\eqref{low_energy_hamiltonian} within the reflection plane
$k_x=-k_y$ onto this eigenspace yields
\bee \label{eqB6}
h({\bf k})_{k_x,+k_y}=
k_{xy}\siy+ (k_z\pm \Delta ) \siz +m \six,\quad   \text{near } (0,0,\mp \Delta ) ,
\ee
where $k_{xy}=(-k_x+k_y)/\sqrt{2}$. 
Because the two Dirac cones in Eq.~\eqref{eqB6} have the same orientation, the total mirror Chern number $C_{x,y}$ changes by two,
when $m \to - m$.

From these observations, we conclude that the total mirror Chern numbers are given by
\begin{eqnarray}
C_{i^0}= 2 \, {\rm sgn}(m) + b_{i}, \quad C_{i,\pm j}= {\rm sgn} \, (m)  + b_{ i,\pm  j} ,
\end{eqnarray}
where $b_{i}$ and $b_{i, \pm j}$ are the mirror Chern numbers of the ``background" bands, i.e.,
those filled bands that are not included in the low-energy description~\eqref{low_energy_hamiltonian}.
We note that there exists the following relation between  $b_{i}$ and $b_{i, \pm j}$
\begin{eqnarray}
b_{i}-b_{ i,\pm  j}=1\ {\rm mod}\ 2. 
\end{eqnarray}
This is because the number of chiral left- (or right-) moving surface modes on the $k_i = 0$ and $k_i = \pm k_j$ high symmetry lines
can only differ by a multiple of two. That is, the surface modes on the $k_i=0$ line are continuously connected to the surface
modes on the $k_i = \pm k_j$ line. The only way how the number of chiral left- (or right-) moving modes can differ on these two high-symmetry lines
is if left- (or right-) moving modes are gapped out pairwise.
%
%

Using the DFT-derived twelve-band tight-binding model and the simplified nine-band model, 
we find that $b_i=0$ and $b_{i,\pm j} =1$.
Hence, $C_{i^0}= 2 \, {\rm sgn}(m)  $ and $C_{i,\pm j}= {\rm sgn} \, (m)   + 1$, 
which is in agreement with the results of Ref.~\cite{TCI_Fu_antiperovskites}.


\section{Landau levels of tilted Dirac cones\label{appendixC}}

{In this section we study the Landau level spectra of type-II Dirac surface states. 
We note that the Landau level structure of tilted Dirac fermions has been studied previously in
the literature~\cite{Kobayashi2007,Goerbig2008,Morinari2009,Goerbig2009,Morinari2010,Kawarabayashi2011,Hatsugai2015,Proskurin2015},
in the context of strained graphene and certain organic conductors, like $\alpha$-(BEDT-TTF)$_2$I$_3$. In this section
we first review the properties of Landau level spectra of titled type-I Dirac cones, and
then extend these results to type-II Dirac surface states.}


\subsection{Effective model approach}

\begin{figure}[!]
\includegraphics[scale=0.8]{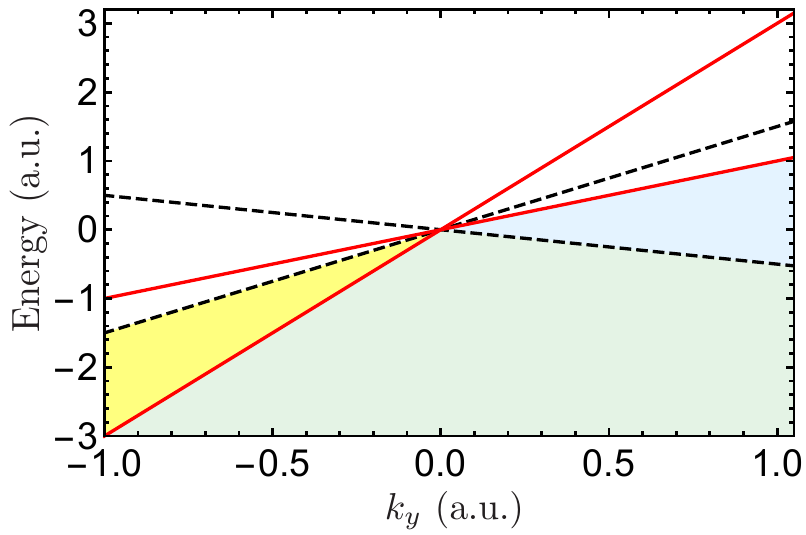}
\caption{\label{Fig:DiracSpectrum} Comparison between a type-I (dashed lines) and a type-II (solid lines) Dirac fermion. 
Shown are the energy dispersions of Eq.~\eqref{toy_ham_C1} with $\eta = 0.5$ (type-I) and $\eta = 2$ (type-II). The valence bands are marked by yellow and light blue for type-I and type-II Dirac fermions, respectively.}
\end{figure}

We start by considering a toy model describing a tilted Dirac cone. The Hamiltonian of this model is given by
\begin{align} \label{toy_ham_C1}
	H_0 = v_F (\eta \pi_y \sigma_0 + \pi_x \sigma_x + \pi_y \sigma_y), 
\end{align}
where $\pi_{i} = \hbar k_{i} + eA_{i}$ denotes the canonical momentum, $\sigma_i$ are the Pauli matrices, and $\eta$ parametrizes the degree of tilting along the $k_y$ direction. 
For $\eta < 1$ ($\eta > 1$) Eq.~\eqref{toy_ham_C1} describes a type-I (type-II) Dirac cone 
with an energy dispersion as shown in Fig.~\ref{Fig:DiracSpectrum}.

The Landau level spectrum of the above Hamiltonian can be obtained in a closed form when $0\leq \eta <1$, which shows that quantized Landau levels  exist for all type-I Dirac cones.
Specifically, if we adopt the Landau gauge $\bm{A} = (0, Bx)$, the spectrum of $H_0$ reads~\cite{Morinari2010}
\begin{align}
	E_n = \sgn{(n)}\sqrt{2eB\hbar v_F^2\abs{n}\lambda^3}, 
\end{align}
where $\lambda = \sqrt{1-\eta^2}$, and $n$ is the Landau level index. 
One intuitive way to understand why Landau level-like spectra still persists for a tilted Dirac cone with $0\leq \eta <1$ is that all constant-energy contours in the momentum space are still closed loops, although with an anisotropic shape~\cite{LiXiao2016:SnTe}. 
As a result, quantized Landau levels can be derived within a semiclassical picture~\cite{Xiao2010}.
Mathematically, the reason for the existence of quantized Landau levels is that the corresponding eigenvalue problem $H_0\Phi(\br) = E\Phi(\br)$ can   be mapped to the problem of a one-dimensional harmonic oscillator, as long as $0 \leq \eta < 1$. Specifically, the eigenstates are governed by the following differential equation~\cite{Morinari2010}, 
\begin{align}
	\brackets{-\dfrac{d^2}{dX^2} + (1-\eta^2)\parenthesis{X+\dfrac{\eta}{1-\eta^2}\varepsilon}^2}\phi(X) \notag\\ = \TwoDMatrix{\dfrac{\varepsilon^2}{1-\eta^2}-1}{i\eta}{i\eta}{\dfrac{\varepsilon^2}{1-\eta^2}+1} \phi(X),\label{Eq:HarmonicOscillator} 
\end{align}
where $\varepsilon = E\ell_B/(\hbar v_F)$, $X = x/\ell_B + k_y\ell_B$, with $k_y$ being the conserved momentum in the Landau gauge, and $\ell_B = \sqrt{\hbar/(eB)}$ is the magnetic length. Because the coefficient of the second term is $1-\eta^2 >0$, harmonic oscillator states are  valid solutions of the above eigenvalue problem, as long as $\eta < 1$.

The above discussion also makes it clear that Landau level-like spectra no longer exists if $\eta >1$, as the coefficient of the second term in the differential equation~\eqref{Eq:HarmonicOscillator} becomes negative. 
Physically, this is because the constant-energy contours now become unbounded (cf. the solid lines in Fig.~\ref{Fig:DiracSpectrum}) in this effective model. 

\begin{figure}[!]
\includegraphics[scale=0.8]{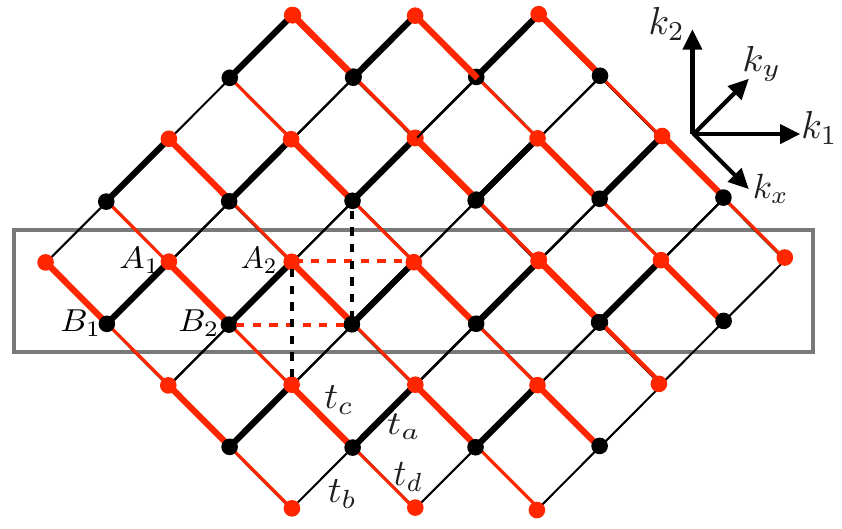}
\caption{\label{Fig:Lattice} 
Illustration of the square lattice, on which the tight-binding model~\eqref{Eq:ToyModel} is defined.
The lattice is bipartite with each unit cell containing two inequivalent atoms labeled by $A$ (red) and $B$ (black). The
second neighbor hopping integral along all dashed lines is $t_1$.
The nearest-neighbor hopping is
specified as $t_a = t_b = t_c = -t$, $t_d = t$.
The gray rectangle marks the ribbon geometry in our calculation, which is periodic along the $k_2$ direction 
and finite along the $k_1$ direction. 
}
\end{figure}

\subsection{Tight-binding model approach}

We now use a tight-binding model to illustrate how the Landau level spectra evolve as the Dirac cone is titled from a type-I cone to a type-II cone.
Specifically, we adopt the following tight-binding model on the square lattice~\cite{Kawarabayashi2011}, 
\begin{align} 
	H = -\sum_{\average{ij}} t_{ij} b_j^\dagger a_{i} + t_1 \sum_{\average{ij}} \left(a_j^\dagger a_i +b_j^\dagger b_i \right), \label{Eq:ToyModel}
\end{align}
where the operator $a_i$ ($b_i)$ annihilates an electron on site $A_i$ ($B_i$). 
The next-nearest-neighbor hopping parameters along the dashed bonds in Fig.~\ref{Fig:Lattice} are given by $t_1$, while the nearest-neighbor hopping amplitude $t_{ij}$ are specified as $t_a = t_b = t_c = -t$, $t_d = t$. 
In the following, we will calculate the spectrum of \eqref{Eq:ToyModel} in a ribbon geometry, as shown by the gray rectangle in Fig.~\ref{Fig:Lattice}, which is periodic along the $k_2$ direction and finite along $k_1$.  

\begin{figure*}[t!]
\includegraphics[scale=0.85]{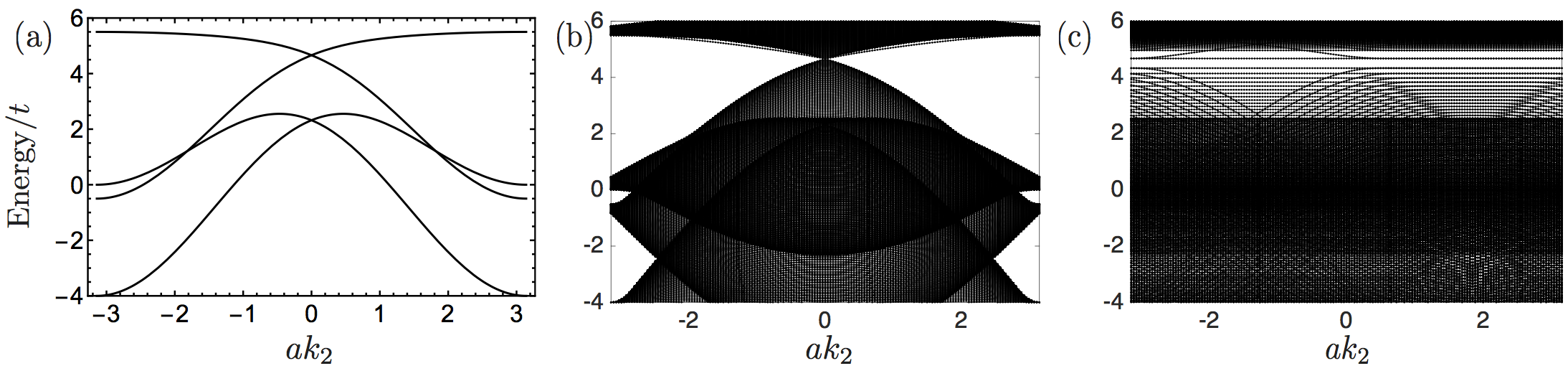}
\caption{\label{Fig:RealModel}  
(a) Bulk energy spectrum of model~\eqref{Eq:RealModel} as a function ok $k_2$ with $k_1 = 0$. 
(b) Energy spectrum of Hamiltonian~\eqref{Eq:RealModel} in a ribbon geometry with the edges along the $k_2$ direction.
(c) Landau level spectrum of Hamiltonian~\eqref{Eq:RealModel} with $\lambda \equiv Ba^2/\phi_0 = 0.05$. 
We note that all dispersive curves in the Landau level structure arise from states localized at the edge of the ribbon. 
}
\end{figure*} 

The energy bands of Hamiltonian~\eqref{Eq:ToyModel} in ribbon geometry are shown in Fig.~\ref{Fig:ToyModel} of the main text. 
The degree of tilting of the two Dirac cones  is controlled by the ratio $t_1/t$. 
Specifically, when $t_1/t = 0$ there is no tilting [Fig.~\ref{Fig:ToyModel}(a)]; 
when $t_1/t$ starts to increase, the two Dirac cones begin to tilt [Fig.~\ref{Fig:ToyModel}(b)]; 
when $t_1/t = 1/2$ the spectrum is nondispersive along the $k_2$ direction [Fig.~\ref{Fig:ToyModel}(c)]; finally, when $t_1/t>1/2$, type-II Dirac cones appear [Fig.~\ref{Fig:ToyModel}(d)]. 

Such a transition from a type-I to a type-II Dirac cone also manifests itself in a drastic change in the Landau level structure, which can be obtained by the Peierls substitution as follows. 
We adopt the Landau gauge and write the vector potential as $\bm{A} = (0, Bx)$, which will attach a phase factor for all hopping processes in the tight-binding model in Eq.~\eqref{Eq:ToyModel}. Specifically, for a hopping from lattice point $(x_i, y_i)$ to $(x_f, y_f)$, the associated phase factor will be $e^{i\theta}$, with 
\begin{align}
  \theta = \dfrac{e}{\hbar}\int_C \bm{A}{\cdot}d\br = \dfrac{(x_f+x_i)(y_f-y_i)}{2\ell_B^2}. 
\end{align} 
Fig.~\ref{Fig:ToyModel} shows the corresponding Landau level spectrum when $\lambda \equiv Ba^2/\phi_0 = 0.01$. 
We can see that when $t_1/t<0.5$, well separated Landau levels exist around the Dirac node, although the level spacing decreases as $t_1$ increases [Figs.~\ref{Fig:ToyModel}(a) and \ref{Fig:ToyModel}(b)]. In contrast, when the transition to a type-II Dirac cone occurs, the spacing of the Landau level around the node becomes extremely small; the Landau levels with finite separation are due to the contributions from other parts of the band structure [Figs.~\ref{Fig:ToyModel}(c) and \ref{Fig:ToyModel}(d)].

{The change in the Landau level structure stems from a change in the Fermi surface topology. In fact, it is generally expected that a large Fermi surface area is associated with dense Landau levels. 
One simple example is type-I Dirac cones with different Fermi velocities: for a given energy, the one with a smaller (larger) Fermi velocity has a larger (smaller) Fermi surface area, which is also associated with a small (large) Landau level spacing. 
One can also gain an intuitive understanding of this from a semiclassical point of view~\cite{Xiao2010,GaoYang}. 
We first note that the Landau levels will occur whenever the semiclassical orbits of electrons in $\bk$-space encloses some critical areas specified by the following condition~\cite{Xiao2010}, 
\begin{align}
  \dfrac{\hat{\bm{B}}}{2} \cdot \oint_{C_m} \bk_c\times d\bk_c = 2\pi\parenthesis{m+\dfrac{1}{2}-\dfrac{\Gamma_{C_m}}{2\pi}}\dfrac{eB}{\hbar}, 
\end{align}
where $C_m$ is the $m$th semiclassical orbit of the electron, and $\Gamma_{C_m}$ is the Berry phase of this energy contour.
We thus see that an additional Landau level will be formed whenever the area of the $\bk$-space semiclassical orbit increases by $2\pi eB/\hbar$. In particular, note that this increment is independent of the Landau level index $m$. 
We can now explain why a large Fermi surface is usually associated with dense Landau levels: a large Fermi surface indicates a semiclassical orbit with a large circumference, and thus a small change in the semiclassical orbit size $\abs{\Delta \bk_c}$ is sufficient to reach the next critical area. 
As a result, as long as the Fermi velocity is not extremely large, we should expect only a small change in the Landau level energy. Therefore, a large Fermi surface area is usually associated with dense Landau levels. 

\subsection{Relation to the $(111)$ surface states}

We now discuss the Landau level structure for the $(111)$ surface states of the antiperovskite Ca$_3$PbO. 
As shown in Fig.~\ref{111}(b) of the main text, the characteristics of the spectrum on the (111) surface is that two type-I Dirac nodes are located at the $\bar{\Gamma}$ point and that their energies are higher than the type-II Dirac nodes away from the $\bar{\Gamma}$ point. 
To describe this surface spectrum we consider the following effective model 
\begin{align}
  H_{\bk}/t = 
  \begin{pmatrix}
    \epsilon_a(\bk) & \gamma_{\bk} & T_{\bk} & 0 \\ 
    \gamma_{\bk}^{\dagger} & \epsilon_b(\bk) & 0 & T_{\bk} \\ 
    T_{\bk} & 0 & \epsilon_c(\bk) & \gamma_{\bk} \\ 
    0 & T_{\bk} & \gamma_{\bk}^{\dagger} & \epsilon_d(\bk)
  \end{pmatrix}, \label{Eq:RealModel}
\end{align}
where $t$ is an overall energy multiplier and 
\begin{align}
  \epsilon_{\alpha}(\bk) = W_{\alpha}[\cos(ak_x) + \cos(ak_y)] + M_{\alpha}, \; \alpha = a, b, c, d .
\end{align}
Here, $a$ is the distance $|A_1B_1|$ in Fig.~\ref{Fig:Lattice}. 
Moreover, we have $\gamma_{\bk} = \sin(ak_x)-i\sin(ak_y)$, and $T_{\bk} = v_2\sin(ak_x)\sin(ak_y)$. 
For the numerical evaluations we choose 
the parameters as $(W_a, M_a) = (2.58, -0.50)$, $(W_b, M_b) = (-0.42, 5.50)$, $(W_c, M_c) = (1.16, 0)$, $(W_d, M_d) = (3.16, -4.00)$, and $v_2 = 0.25$, respectively. 
The energy spectrum of this model at $k_1 = 0$ is shown in Fig.~\ref{Fig:RealModel}(a), which captures the Dirac features of the $(111)$ surface state [cf. Fig.~\ref{111}(b) in the main text]. {We note that this effective model possesses a $C_3$ rotation symmetry, instead of the $C_4$ rotation symmetry in the actual (111) surface of the antiperovskites; hence, there are four type-II Dirac nodes, instead of six. }


In order to calculate the Landau level spectrum of this model, we 
assume that it is defined on the square lattice shown in Fig.~\ref{Fig:Lattice}, where each site now hosts four orbitals $(A_{\bk}, B_{\bk}, C_{\bk}, D_{\bk})$, which constitutes the basis of the Hamiltonian $H_{\bk}$, Eq.~\eqref{Eq:RealModel}.
For convenience, we also make a coordinate transformation, namely $k_1 = k_x+k_y$, and $k_2 = k_x-k_y$. We then keep the system periodic along the $k_2$ direction, while finite along the $k_1$ direction. In particular, we only retain the lattice points marked by the gray rectangle in Fig.~\ref{Fig:Lattice}. 
The energy spectrum of such a ribbon geometry is shown in Fig.~\ref{Fig:RealModel}(b). 
The Landau level spectrum of~\eqref{Eq:RealModel} for $\phi/\phi_0 = 0.05$ is shown in Fig.~\ref{Fig:RealModel}(c). 
The type-I and type-II Dirac nodes exhibit distinguishable physical features. Near the type-I Dirac node at $E/t\sim 4.5$, Landau levels are well separated since only a single Fermi surface appears near the node. Near the type-II Dirac nodes and the second type-I Dirac node at $E/t\sim 1.5$, on the other hand, the spacing of the Landau levels is close to zero due to the complexity of the Fermi surface structures.


\section{Surface states for the other termination}
\label{appendixD}

\begin{figure}[t]
\includegraphics[clip,width=0.99\columnwidth]{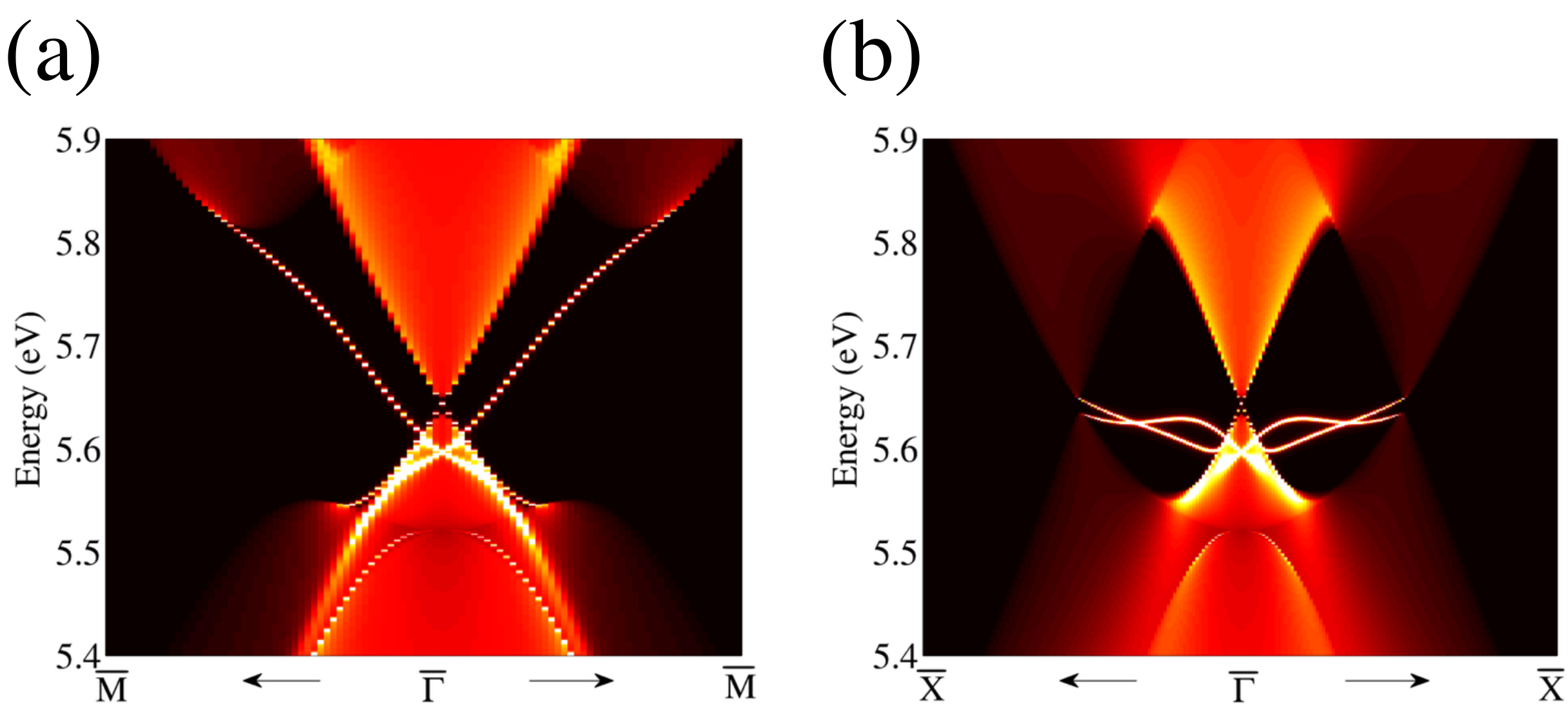}
\caption{\label{Fig:001bottom} \label{mFigS4}
Surface density of states along the high symmetry lines (a) $\bar{\Gamma} \to \bar{X}$ and (b) $\bar{\Gamma} \to \bar{M}$ of the (001) surface BZ for the Ca-O termination [lower right panel of Fig.~\ref{mFig1}(b)].}
\end{figure}

For completeness we show in this section the surface states for the other surface terminations.

Figure~\ref{Fig:001bottom} displays the Dirac states on the (001) surface for the Ca-O termination.
The surface density of states is plotted along the
high-symmetry lines $\bar{\Gamma} \to \bar{X}$ and $\bar{\Gamma} \to \bar{M}$, corresponding to the $k_x=0$ 
and $k_x = k_y$ mirror lines, respectively. 
Both in Figs.~\ref{Fig:001bottom}(a) and~\ref{Fig:001bottom}(b) two chiral left- and right-moving surface states are clearly visible,
which connect valence with conduction bands. This is in agreement with the mirror Chern numbers $C_{x^0}$ and
$C_{x,y}$ which take the value 2.
In Fig.~\ref{Fig:001bottom}(a) there is in addition a trivial surface state which intersects with one of the left- (right-)moving chiral modes.

Figure~\ref{Fig:110bottom} shows the Dirac states on the (011) surface for the Ca termination. 
The surface density of states is plotted along the high symmetry lines $\bar{\Gamma} \to \bar{X}$
and $\bar{\Gamma} \to \bar{Y}$, which corresponds to the $k_x=0$ and $k_y = k_z$ mirror lines, respectively.
Since $C_{x^0} = C_{y,z}=2$ there appear two left- and two right-moving chiral modes. 
In Fig.~\ref{Fig:110bottom}(b) the chiral modes form a type-II Dirac corssing.

In closing we note that within our  twelve-band tight-binding description the (111) surface spectrum with O termination is identical to the one with Ca-Pb termination. This is because our tight-binding model does not include oxygen orbitals, since they are far away in energy from $E_F$. }


\begin{figure}[t]
\includegraphics[clip,width=0.99\columnwidth]{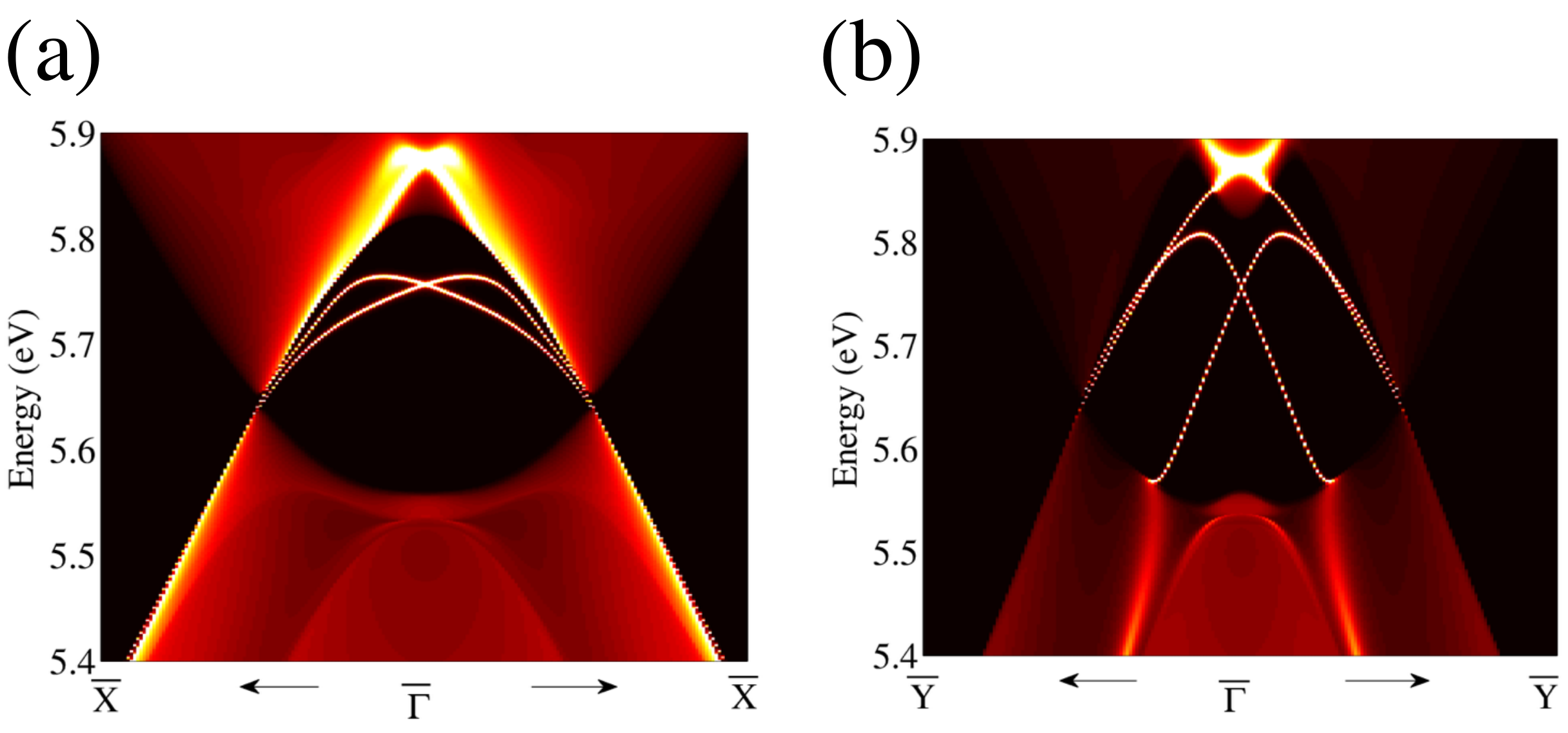}
\caption{\label{Fig:110bottom} Surface density of states along the high symmetry lines (a) $\bar{\Gamma} \to \bar{X}$ and (b) $\bar{\Gamma} \to \bar{Y}$ of the (011) surface BZ for the Ca termination.}
\end{figure}

\clearpage

\end{document}